# Ultrafast Photoinduced Band Splitting and Carrier Dynamics in Chiral Tellurium Nanosheets

Giriraj Jnawali*[1], Yuan Xiang[2], Samuel M. Linser[1], Iraj Abbasian Shojaei[1], Ruoxing Wang[3], Gang Qiu[4], Chao Lian[5], Bryan M. Wong[5], Wu Wenzhuo[3], Peide D. Ye[4], Yongsheng Leng[2], Howard E. Jackson[1] and Leigh M. Smith*[1]

[1] Department of Physics and Astronomy, University of Cincinnati, Cincinnati, OH 45221, USA
[2] Department of Mechanical & Aerospace Engineering, The George Washington University, Washington, D.C. 20052, USA
[3] School of Industrial Engineering, Purdue University, West Lafayette, IN 47907, USA
[4] School of Electrical and Computer Engineering, Purdue University, West Lafayette, IN 47907, USA
[5] Department of Chemical & Environmental Engineering, Materials Science & Engineering Program, University of California, Riverside, Riverside, CA 92521, USA

*Corresponding authors

## Abstract

Trigonal tellurium (Te) is a chiral semiconductor that lacks both mirror and inversion symmetries, resulting in complex band structures with Weyl crossings and unique spin textures. Detailed time-resolved polarized reflectance spectroscopy is used to investigate its band structure and carrier dynamics. The polarized transient spectra reveal optical transitions between the uppermost spin-split $H_4$ and $H_5$ and the degenerate $H_6$ valence bands (VB) and the lowest degenerate $H_6$ conduction band (CB) as well as a higher energy transition at the L-point. Surprisingly, the degeneracy of the $H_6$ CB (a proposed Weyl node) is lifted and the spin-split VB gap is reduced upon photoexcitation before relaxing to equilibrium as the carriers decay. Using ab initio density functional theory (DFT) calculations we conclude that the dynamic band structure is caused by a photoinduced shear strain in the Te film that breaks the screw symmetry of the crystal. The band-edge anisotropy is also reflected in the hot carrier decay rate, which is a factor of two slower along c-axis than perpendicular to it. The majority of photoexcited carriers near the band-edge are seen to recombine within 30 ps while higher lying transitions observed near 1.2 eV appear to have substantially longer lifetimes, potentially due to contributions of intervalley processes in the recombination rate. These new findings shed light on the strong correlation between photoinduced carriers and electronic structure in anisotropic crystals, which opens a potential pathway for designing novel Te-based devices that take advantage of the topological structures as well as strong spin-related properties.

# Introduction

The isolation of graphene[1] has stimulated extensive research in van-der Waals (vdWs) materials revealing new physics, as well as offering building blocks for novel electronic devices.[2-4] Beyond graphene and the transition metal dichalcogenides (TMDs), a variety of exotic materials such as topological insulators (TIs)[5] and Weyl semimetals (WSMs)[6] have been recently discovered, which display unique electronic structure and chiral spin texture on the Fermi surface. Most of these materials involve heavy elements such as Bi, Sb, Te, Se, etc., suggesting that strong spin-orbit interaction (SOI) is the key to the complex electronic structure in these materials. Elemental tellurium (Te) has strong interest due to its chiral nature and unique spin texture and topological features in the band structure.[7-11] Both 2D Te nanostructures and high mobility field-effect devices have been demonstrated.[12]

Trigonal Te is a 1D vdWs crystal where 3-fold helices of covalently bonded Te atoms are weakly coupled in a hexagonal close-packed array.[13, 14] Each helix is parallel to the c-axis (see Fig. 1a), and so crystals display either right-handed (space group: $P3_121 - D_3^4$) or left-handed (space group: $P3_221 - D_3^6$) symmetries.[15] Gyrotropic properties have been observed in Te such as a strong optical rotatory power,[16] the circular photogalvanic effect,[17, 18] current-induced magnetization,[19] etc. This makes trigonal Te a unique material for polarization optics, multiferroics, and spintronics. The lack of mirror and inversion symmetries result in unique radial spin texture in Te band structure,[7-9] in contrast to Rashba systems[20] or topological materials.[21] The spin degeneracy of the uppermost valence bands (VBs) at the H- or H'-points in the Brillouin zone, is lifted in momentum space due to the SOI and the breaking of inversion symmetry while the lowest conduction band (CB) is doubly spin-degenerate and protected by the three-fold screw symmetry of the helices.[7-10] Trigonal Te is insulating at ambient pressure but transforms to a metallic phase under hydrostatic pressure.[22, 23] The band structure of Te exhibits a topological electronic structure with a number of Weyl nodes near the H-point, particularly the spin-degenerate $H_6$ CB minimum.[9, 10] The topological phase transition from a trivial semiconductor to a Weyl semimetal (WSM) is predicted under applied external pressure when the spin-polarized uppermost VBs and CB are inverted across the band gap.[8, 24] Since a strong piezoelectric effect has been reported in bulk and nanostructured Te,[25-27] strain can also be generated in response to an applied electric field and *vice versa*. The response of the electronic structure to laser excitation could enable manipulation and control of topological phases in Te. The unique chiral electronic structure offers the ability to control the electronic charge and spin degrees of freedom in the absence of magnetic field for applications in spintronics.

While ground-state optical absorption with light polarized parallel or perpendicular to the c-axis has been carried out,[15, 28-32] very little is known about carrier dynamics or higher lying states. The ground-state band-edge absorption anisotropy is explained by the symmetry properties of the Te crystal.[32, 33] However, at higher energies the Te band structure exhibits a large number of multi-valley nested degenerate band structures that results in high thermoelectric efficiencies.[34] Probing carrier dynamics in such a wide energy region provides a unique opportunity to investigate inter- and intra-valley carrier decay processes in addition to the fundamental recombination lifetime.[35]

Polarized optical transitions are studied in a Te nanosheet at 10 K and 300 K by exciting with a 1.51 eV pump pulse and probing with a tunable mid-infrared probe pulse polarized parallel and perpendicular to the c-axis. Polarized optical transitions and carrier dynamics at the band-edge and higher lying states are probed over nanoseconds and analyzed with the support of numerical calculations. Transient anisotropic band modifications observed in the measurements are proposed to be caused by a photoinduced piezoelectric shear strain generated in the Te film and explicitly calculated by ab initio density functional theory (DFT) band structure calculations. Energy and polarization dependent kinetics are evaluated to determine the dominant relaxation channels of photoexcited carriers in Te.

## Results

**Sample characteristics.** The Te nanostructures are chemically synthesized and dispersed onto a Si/SiO$_2$ substrate (see Methods).[12] The covalently bonded Te chains are parallel to the c-axis of the hexagonal crystal while the basal ab–plane is perpendicular to the chain (Fig. 1a). The c-axis lies in the nanosheet plane enabling polarized optical measurements. Figure 1b shows AFM images of the nanostructures and their orientation with respect to the c-axis. The thickness varies between 20 nm to 30 nm (see Fig. 1b). The surface morphology displays an atomically smooth surface, without noticeable oxidation.

The sample orientations were determined by polarized micro-Raman spectroscopy (see Methods) with the laser polarized parallel ($E \parallel c$) or perpendicular ($E \perp c$) to the c-axis. With $E \perp c$, we observe three sharp Raman peaks at 119 cm$^{-1}$, 90 cm$^{-1}$, and 138.5 cm$^{-1}$ corresponding to $A_1$ (symmetric intra-chain breathing in the basal plane), and the doubly degenerate $E(1)$ (rigid chain rotation in the basal plane), and $E(2)$ (asymmetric stretching along the chain or the c-axis) modes, as indicated in Fig. 1c.[36] The weak second order harmonic $E'$ at $266 \pm 1$ cm$^{-1}$ is also seen. Logarithmic plots of the data (see inset of Fig. 1c) confirm that the $E(1)$ mode is symmetry forbidden for $E \parallel c$.

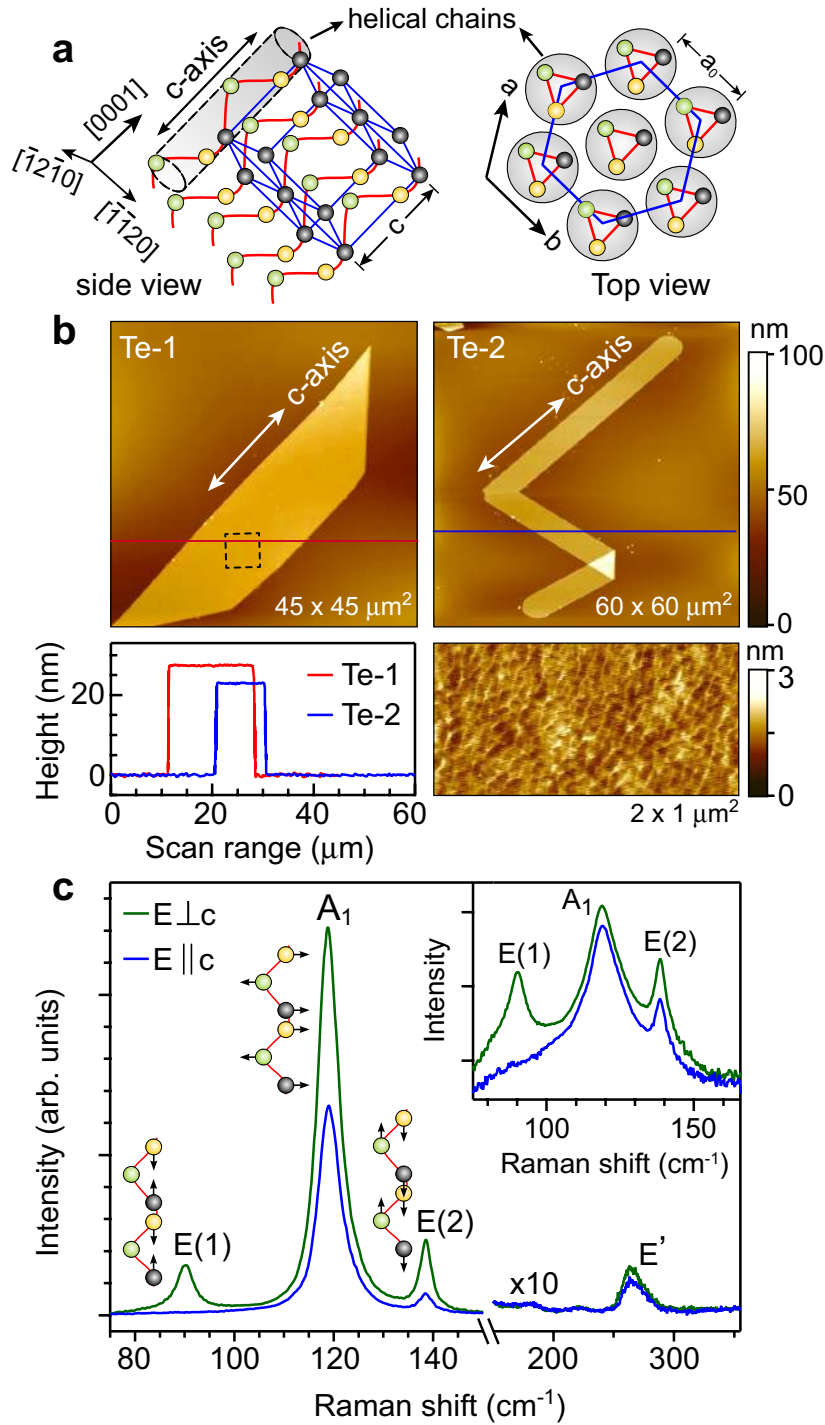

**Figure 1 | Structure and morphology of Te nanosheets.** (**a**) Crystal structure of trigonal Te (left image) in which spheres of different colors represent Te atoms within the helical chains (shaded region) arranged in a hexagonal array along the growth direction, i.e., c-axis. Pattern with blue lines shows a hexagonal unit cell along [0001]. The right image is the top (surface) view of the crystal from [0001] direction, showing ab–plane (basal plane) and surface lattice parameter ($a_0$). (**b**) AFM topographies of the Te nanosheets used for optical measurements (symbolized by Te-1 and Te-2) and their cross section profiles (red and blue lines). Close-up height scan on Te-1

(dotted rectangle) is shown on bottom right, which exhibits relatively smooth surface with an average corrugation of < 2 nm. (**c**) Representative Raman spectra acquired on these samples using a 633 nm laser excitation source with two perpendicularly polarized configurations with respect to the c-axis, i.e., $E \parallel c$ and $E \perp c$. Three prominent optical modes, as indicated by $A_1$, $E(1)$, and $E(2)$ modes, respectively, are detected with $E \perp c$ while only $A_1$ and $E(2)$ modes are clearly visible with $E \parallel c$. Inset shows a log scale plotting of the data, showing that the $E(1)$ mode is symmetry forbidden for $E \parallel c$. A weak second order harmonic of $E$-mode is seen in the spectra and indicated by the symbol $E'$.

**Polarized transient reflectance (TR) spectroscopy.** Polarized TR measurements are performed using pump-probe methods (see Methods and Supplementary Note 2). The change in the reflectance of the polarized tunable probe beam with the 1.5 eV pump pulse present, $\Delta R(E, t) = (R_{pump-on} - R_{pump-off})$, is measured as a function of both energy ($E$) and delay ($t$). The $\Delta R(E, t)$ data are normalized by the polarized probe reflectance $R_0(E)$ (pump off) at all probe energies $E$, and recorded as $\Delta R/R_0$. Figure 2a displays a false color map of the TR spectra, $\Delta R/R_0$ measured over $0.3 - 1.2$ eV energies for the probe polarized parallel ($E \parallel c$) and perpendicular ($E \perp c$) to the c-axis. The spectra show derivative-like anti-symmetric features in both the low-energy ($\sim 0.3 - 0.5$ eV) and high-energy ($\sim 1 - 1.2$ eV) regions and a relatively weaker and broader signal near $\sim 0.6 - 0.8$ eV, indicating a series of interband optical transitions. The $E \parallel c$ spectra exhibit an additional anti-symmetric feature near $\sim 0.45$, which is not present for $E \perp c$, suggesting polarization-sensitive transient optical response in Te samples. We first quantify each of these features and compare to band-to-band optical transitions from band-structure calculations.

**Ground-state optical transitions.** The TR signal near the band-edge decays almost completely within first 30 ps (Fig. 2a) with a weak residual signal persisting over a longer time. In Fig. 2b we display normalized spectra at 100 ps delay for each polarization. The derivative-like features of the transient signal at low energies $(0.34 - 0.45\, \text{eV})$ exhibit a clear difference for the two polarizations while high-energy features $(0.9 - 1.2\, \text{eV})$ are nearly identical. These features persist over long time scales, suggesting their origin is due to weak perturbation of the dielectric response caused by photoexcitation of electrons and holes. The energies of optical transitions are accurately determined by using a derivative Lorentzian lineshape,[37]

$$\frac{\Delta R}{R_0}(E) \simeq \sum_{j=1}^{n} Re\left[A_j e^{i\varphi_j}(E - E_j + i\Gamma_j)^{-2}\right], \qquad (1)$$

where $n$ represents the number of interband transitions involved, $E$ is the photon energy of the probe beam, and $A_j$, $\varphi_j$, $E_j$, and $\Gamma_j$ are the amplitude, phase, transition energy, and the energy broadening parameter of the $j^{th}$ feature, respectively. Such a form cannot provide physical insights, but provides an efficient way of accurately determining the transition energies. The best-fit lineshapes to spectra are shown with dashed-lines in Fig. 2b. The spectra with $E \perp c$ can be fitted by three resonances ($n = 3$) while the spectra with $E \parallel c$ can be fitted by four resonances ($n = 4$). Moduli (absorption profiles) of individual resonances obtained from Eq. (1), are shown with minor vertical shifts for clarity. The values of the transition energies are indicated by vertical dotted lines.

The transition energies determined above are compared with the detailed band structure around the H and H' high symmetry points of the bulk Brillouin zone of Te,[7, 9, 38] (see Fig. 2c and inset). The five bands closest to the Fermi level probed in the TR experiment include three successive VBs and one CB minimum. The uppermost two VBs ($H_4$ and $H_5$) are spin-polarized (non-degenerate) while the lower lying VB ($H_6$ or $H_6^{VB}$) and the lowest CB ($H_6$ or $H_6^{CB}$) are Rashba-like 2-fold spin-degenerate and protected by the screw symmetry of the helices.[7-9] The $H_6^{CB}$ CB minimum has been proposed to be a Weyl node crossing.[9] The intrinsic large SOI causes a camelback feature in the $H_4$ VB along $H-K$ direction, showing a weakly indirect band gap in Te. In our experiments, the 1.5 eV pump photon energy generates hot electrons and holes and these carriers scatter to band minima through inter- and intra-band processes. The time-delayed mid-IR probe pulse interrogates the carrier occupation of each band by monitoring the change in the reflectivity at specific interband transition energies (see vertical arrows). The energies of these three optical transitions nearly match the energy of each band as predicted by theory (Fig. 2c).

The lowest $H_4 \to H_6^{CB}$ transition with $E \perp c$ at 100 ps is $E_1^\perp = 0.363 \pm 0.002$ eV, which is almost identical with the transition with $E \parallel c$, $E_1^\parallel = 0.366 \pm 0.002$ eV. Both transitions appear to be direct and are nearly identical in energy, though a one-third weaker response is observed with $E \parallel c$, caused by the higher reflectivity $R_0$ of the sample with $E \parallel c$ as compared to $E \perp c$.[28] We do not see any evidence of a forbidden or indirect transition with $E \parallel c$ as discussed in previous work.[15, 28] The transition $E_1^\perp$ is ~30 meV larger than the previous optical absorption results of the bulk band gap of Te.[15, 28, 39] The higher absorption-edge is caused by the Moss-Burstein shift of the Fermi energy below the maximum at $H_4$ in our naturally $p$-doped samples (see Fig. 2c). Hall measurements[12] in similar nanosheets have indicated a free hole density of ~5×10$^{18}$ cm$^{-3}$. The optical transition $H_5 \to H_6^{CB}$ is observable only with $E \parallel c$, consistent with the expected dipole allowed transition between these two states.[40] The transition energy at longer delays is $E_2^\parallel = 0.448 \pm 0.005$ eV, so we conclude that the energy separation between the Fermi level in the $H_4$ VB and the $H_5$ VB is $E_2^\parallel - E_1^\perp = 0.085$ eV. We determine a 30 meV Fermi energy within the $H_4$ band, indicating the gap between the $H_4$ and $H_5$ band-edges to be $\Delta E = 0.115$ eV, which coincides with the 11-micron wavelength hole absorption band, previously measured in a semi-insulating bulk single crystal of Te.[29, 41]

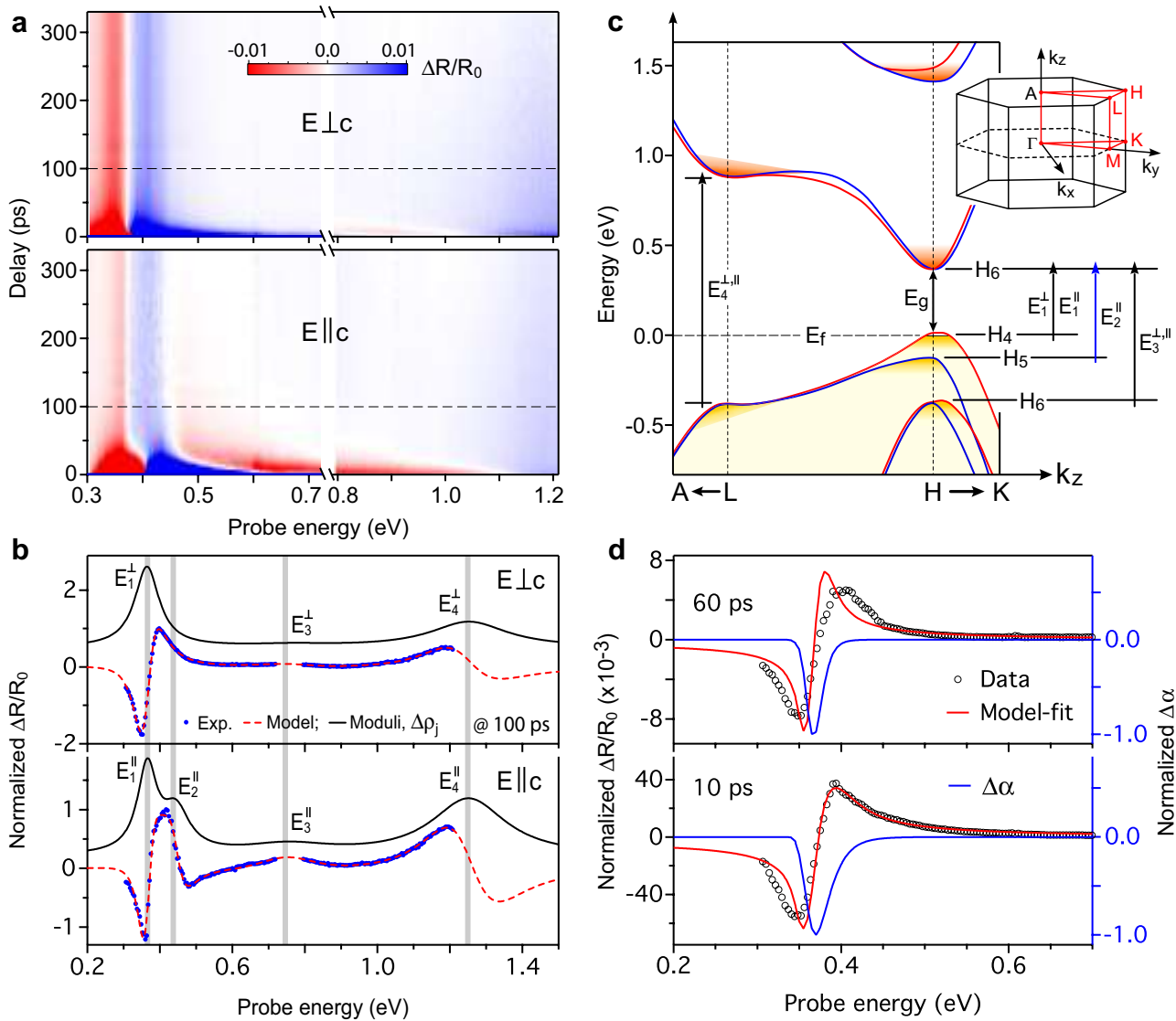

**Figure 2 | Polarized spectral map and modeling of the TR response of a Te nanosheet.** (**a**) Two-dimensional false color maps of the pump-probe delay dependent polarized TR spectra over extended probe energies, showing strong features with distinctive dynamics at different energies. (**b**) TR spectra (blue circles) acquired at 100 ps time delay with two orthogonal probe polarizations. Each spectrum is fitted with a simple model described in the main text (red dashed lines). Calculated moduli $\Delta\rho_j$ of each fit are also plotted with a vertical offset (black lines). Both the model fits and the moduli show transient features corresponding to the series of optical transitions in the Te bulk band structure. (**c**) Schematic band structure of bulk Te and possible optical transitions around the high symmetry H and L points of the Brillouin zone near the band-edge region. Observed transition energies in the TR spectra at different band-edge regions are indicated. On the right top, the first Brillouin zone of Te and various high symmetry points are indicated. (**d**) Transient spectra (black open circles) around the fundamental gap region measured with the $E \perp c$ configuration of the probe pulse at two different time delays. Each spectrum is well reproduced using a band-filling model described in the main text (red lines). The calculated change in absorption

Δα, normalized by the peak values for clarity, at respective delay times are also superimposed with scales on the right (blue lines). Overall, the model captures the main features of the spectra and extracts important band parameters such as band gap and Fermi energy.

The energy and polarization dependence of the optical transitions is strongly consistent with band calculations. In addition to the direct transitions at the band-edge, a higher energy transition near $E_3^{\parallel,\perp} = 0.75 \pm 0.005$ eV is observed which is weakly sensitive to probe polarization. The one order of magnitude weaker TR signal at this energy suggests that this transition might be indirect in **k**-space. The valence band $H_6^{VB}$ (Fig. 2c) is doubly spin-degenerate with one spin-state showing a maximum outside $\mathbf{k_z} = 0$ at the H-point before extending to the Weyl node.[7, 9] This indirect transition is likely between $H_6^{VB}$ and $H_6^{CB}$. The transition at $E_3^{\parallel,\perp}$ falls right in the gap of our photon source, $0.72 - 0.8$ eV, which adds a small uncertainty in estimating the transition energy. A higher lying transition at $E_4^{\parallel,\perp} \sim 1.25 \pm 0.005$ eV, is nearly four-times the gap. The TR response of this transition is a factor of two larger than the $E_3^{\parallel,\perp}$ response but much weaker than the band-edge transitions. Due to the probe pulse energy range, only one-half of the derivative-like line shape is accessible, causing uncertainties in extracting the precise energy and polarization anisotropy of this transition. This higher energy transition cannot result from higher lying valence or conduction bands at the H-point. Based on the band-structure of Te near high symmetry points in the Brillouin zone,[7, 9] we find only one vertical transition possible at the L-point that matches with $E_4^{\parallel,\perp}$.

**Estimation of band-edge parameters.** A band-filling analysis is used to model the TR spectra at the band-edge. Our samples are highly p-type,[12] and so the lowest observed optical transition is between the Fermi level below the VB edge to the bottom of the CB. The TR lineshapes result from changes to the complex index of refraction caused by the photoexcited electrons and holes. The additional holes cause a slight downshift of the Fermi energy at the $H_4$ VB, which changes the absorption onset for optical transitions between the $H_4$ VB hole gas to the $H_6^{CB}$ CB. Using Kramers-Kronig analysis, the calculated change in the absorption coefficient (Supplementary Note 5) is transformed to a change in the real part of the index of refraction. The free parameters in this calculation are the density of the doped hole gas in $H_4$ band, the density and temperature of the photoexcited carriers (electron-hole pairs), and the fundamental gap energy $E_{g1}(H_4 \rightarrow H_6^{CB})$. The lineshapes are calculated assuming parabolic electron and hole bands with effective masses $0.6 \cdot m_e$ and $0.4 \cdot m_e$, respectively.[7] The calculated Fermi energy is smaller because of the substantial valley degeneracies of $H_4$ and $H_6^{CB}$ bands, 12 and 24, respectively. Figure 2d shows TR spectra (black open circles) and model fits (red lines) near the band-edge measured with $E \perp c$ at two different delay times. The blue lines overlapped on each spectrum with scales on the right show a decrease in the absorption (photoinduced bleaching) due to the pump excitation. Best fits show a doped hole density of $N_d \sim 2 \times 10^{18}$ cm$^{-3}$, with Fermi energy of $E_f \sim 30$ meV, and a fundamental gap between $H_4$ and $H_6^{CB}$ of $E_g \sim 0.32 \pm 0.005$ eV. Our estimated doping agrees well with those obtained by Hall measurements on similar nanosheets.[12, 42] The photoexcited carrier density at the band-edge reaches to $\Delta N_{eh} \sim 1 \times 10^{18}$ cm$^{-3}$ at a delay time of 10 ps. The zero crossing of the lineshape occurs at the minimum of the absorption change $\Delta\alpha$ (right scale in Fig. 2d) at 0.368 eV, which matches closely with $E_1^\perp$ and provides a consistency check of the Aspnes form fit (eq. 1). The estimated fundamental gap is consistent with previous low temperature optical measurements on moderately doped bulk samples.[15, 28, 39] The band parameters obtained from the model fittings are tabulated in Supplementary Table 1,2.

**Dynamic evolution of optical transitions.** Figure 3a,b shows selected TR spectra at different delay times for orthogonal polarizations near the band-edge. Apart from amplitude, the lineshape does not change for $E \perp c$, likely caused by the relatively weak pump excitation. The spectra with $E \parallel c$, however, display changes caused by a band shift. Due to the complex Te VB structure, accurate modeling of the $E \parallel c$ spectra is not trivial; by assuming minimal broadening we analyze these spectra by fitting with Eq. (1). Aspnes et al. have shown that increasing the broadening by a factor four or amplitude by a factor of 100 have minimal impact (< 3 meV) on the determination of the transition energy.[43] As described earlier, the spectra with $E \perp c$ fit well with a single transition energy ($n = 1$) while the spectra with $E \parallel c$ fit well with a two transition energies ($n = 2$), close to each other but phase-shifted by 180° (dashed red lines). Corresponding moduli (absorption bands) of the spectra obtained from the fits are shown in Fig. 3c,d. The time evolution of the peak position exhibits a shift in the transition energies for both polarizations and are plotted as a function of delay in Fig. 3e. The lowest transition energy with $E \perp c$ remains steady at the ground-state value of $E_1^\perp = 0.36$ eV except a minor down-shift ($\sim 3$ meV) within 2 ps. Such an abrupt down-shift of band-gap is attributed to band-gap renormalization, which results from many-body effects arising from the hot electron-hole plasma.[44] The lowest energy transition with $E \parallel c$, however, is shifted by $\sim 20$ meV higher than $E_1^\perp$ near time zero. The higher energy $E_2^\parallel$ transition between $H_5 \to H_6^{CB}$ with $E \parallel c$ (dipole forbidden with $E \perp c$), is red-shifted by $\sim 10$ meV near time zero. Interestingly, both of these transitions with $E \parallel c$ gradually relax back to their respective late time energies within $\sim 30$ ps. The transient recovery is shown by plotting the delay dependence of the energy difference near the first and second band-edge regions, i.e., $E_1^\parallel - E_1^\perp$ and $E_2^\parallel - E_1^\perp$, respectively, as well as spin-split VB gap, i.e., $E_2^\parallel - E_1^\parallel$ (inset of Fig. 4e). The maximum shift $E_1^\parallel - E_1^\perp$ near time zero is $\sim 20$ meV while the $E_2^\parallel - E_1^\parallel$ is reduced by $\sim 30$ meV. An exact evaluation of the spin-split gap is hindered by the lack of well-resolved transitions from $H_5$ to the closely spaced $H_6^{CB}$ lifted bands. The strong line broadening of the $E_2^{\parallel}$ TR spectra as compared to the $E_1^{\parallel}$ transition at early times can be seen in the green and red dashed fits in Fig. 3d. The shifts in both transitions gradually return to steady states as the majority of photoexcited carriers relax to the ground-state within first ~ 30 ps.

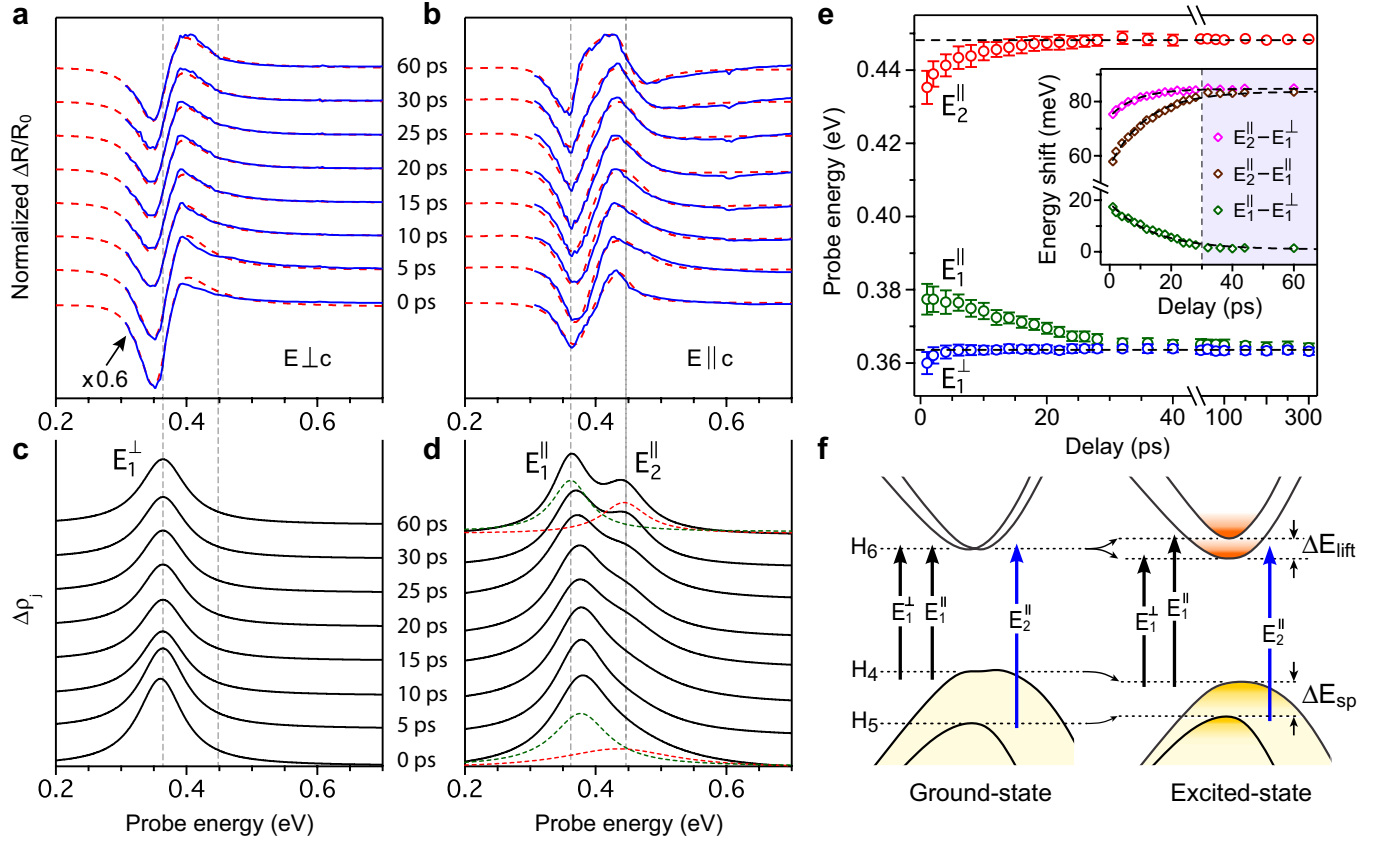

**Figure 3 | Transient response of polarized optical transitions in Te nanosheet.** (**a**,**b**) A series of selected TR spectra (blue lines) recorded at different pump-probe delay times for both $E \perp c$ and $E \parallel c$ polarizations. Dashed red lines are model fits of each corresponding spectrum. (**c**,**d**) Moduli of each corresponding fit for both polarizations. Vertical dashed grey lines are guides to the eye to indicate the systematic shift of peak position. (**e**) Transition energies $E_1^{\perp}$, $E_1^{\parallel}$, and $E_2^{\parallel}$ extracted from the fits and plotted as a function of delay for both polarizations. Error bars are estimated from the least square fitting procedure. While the $E_1^{\perp}$ remains mostly stable over a long delay, the $E_1^{\parallel}$, and $E_2^{\parallel}$ undergo gradual blue and redshifts as delay progresses before they remain steady after ~ 30 ps. The inset shows the time evolution of anisotropic spectral shifts, as quantified by $E_1^{\parallel} - E_1^{\perp}$ and $E_2^{\parallel} - E_1^{\perp}$ for fundamental and higher energy band-gap anisotropy and $E_2^{\parallel} - E_1^{\parallel}$ for the gap between uppermost VBs excluding the doping induced Fermi energy shift. Superimposed dashed lines on each curve are exponential fits for a guide to the eye. The shaded region is shown to indicate the static region. (**f**) Ground-state and photoinduced band structure (note exaggerated energy scales) around the H-point near the band-gap, showing photoinduced lifting $\Delta E_{lift}$ of $H_6^{CB}$ band and narrowing of the gap $\Delta E_{sp}$ between the uppermost VBs. The involved optical transitions are indicated by vertical arrows.

The physical origin of this dynamic shift in the first 30 ps is complex. Because the shifts are both positive and negative for different transitions, it is not possible to explain the shifts due to ultrafast processes including excitonic, many body or band filling effects. Excitonic interactions are ruled out due to the very low exciton binding energy ($E_b \sim 0.5$ meV $\ll E_g$) in bulk Te.[45] An ultrafast band-gap collapse upon high density ($> 1$ % of valence electrons or $\sim 10^{21}$ cm$^{-3}$) excitation is predicted in Te,[46] which is not possible here ($n_{eh} \sim 10^{18}$ cm$^{-3}$). The only remaining possibility is a transient change in the lattice structure of Te causing a dynamic shift in the band-edges. Recent theoretical studies have predicted strain-induced band modifications in a Te crystal.[8, 24] Under uniaxial or hydrostatic pressure, all CBs and VBs edges move vertically, either decreasing (compressive strain) or increasing (tensile strain) the gap, but the degeneracy of the $H_6^{CB}$ band is protected. Shear strain, while affecting all the bands, also lifts the $H_6^{CB}$ band degeneracy by breaking the screw symmetry of the Te helix.[8] We therefore rule out uniaxial or hydrostatic stress effects because the gap or $E_1^{\perp}$ transition with $E \perp c$ remains unchanged upon photoexcitation. Moreover, there is no change in the band-edge shift for the pump pulse aligned parallel or perpendicular to c-axis, in contrast to the uniaxial strain effect in anisotropic Te.[8, 47] Thus shear strain must contribute dominantly in our experiment. Furthermore, an induced shear strain should change the spin-split VB gap due a change in SOI;[48] this is observed in our experiment. We propose that these dynamic shifts in the first 30 ps arise due to lifting of the $H_6^{CB}$ spin degeneracy and the decrease of the spin-split VB gap by photoinduced shear strain generation in Te helices, as shown schematically in Fig. 3f. The lifting of the $H_6^{CB}$ band degeneracy $\Delta E_{lift}$ increases the energy of the $E_1^{\parallel}$ and decreases the $E_2^{\parallel}$ transitions as well as the spin-split VB gap $\Delta E_{sp}$ consistent with experiments. These strain dependent shifts observed in $E \parallel c$ spectra relax within the first $\sim 30$ ps, suggesting strong correlation between the density of photoexcited carriers and electronic structure.

To support our proposed strain-induced band modifications, we have used density functional theory (DFT) to calculate the Te band structure under different strains (see Methods). Under zero or uniaxial/hydrostatic strains, the band structure shows identical features as described in both early[8] and recent[24] studies (Supplementary Note 1). As expected, VB and CB edges move vertically but preserve the band degeneracy of $H_6^{CB}$. Shear strains applied along $[\bar{1}2\bar{1}0](x)$ or $[10\bar{1}0](y)$ directions, on the two (0001) surfaces, however, cause the two fold degeneracy of $H_6^{CB}$ to lift even for a small 0.75 % shear strain, forming two nested energy valleys (Fig. 4a). Furthermore, both the splitting $\Delta E_{lift}^{calc}$ of the $H_6^{CB}$ band increases, and the spin-split VB gap $\Delta E_{sp}^{calc}$ decreases systematically with strain (Fig. 4a,b). The fundamental gap is only weakly affected by shear strain. The calculated degeneracy lifting $\Delta E_{lift}^{calc}$ of the $H_6^{CB}$ under 3 % strain along the $x$-direction is 24 meV, and along the $y$-direction is 30 meV. Both values are higher than our experimentally estimated value of $\Delta E_{lift} \sim 17$ meV at $t = 0$ ps. The calculated spin-split VB gap $\Delta E_{sp}^{calc}$ is hardly affected by $y$-direction strain but noticeably decreases for $x$-direction strain. This suggests that the $[\bar{1}2\bar{1}0](x)$ shear strain is dominant. These shear strain calculations capture all the features of the dynamic band modifications we observe in our measurements. However, the magnitude of $\sim 2 - 3$ % strain required to match our measurements is somewhat large. Considering the elastic strain field in the Te flake induced by laser excitation is a complex three-dimensional strain field (the calculations assume uniform shear strain), we anticipate that if the equivalent von Mises shear strain induces the same shear lifting of $H_6^{CB}$ degeneracy, the corresponding shear strain along the $[\bar{1}2\bar{1}0]$ or $[10\bar{1}0]$ direction should be smaller than the estimated 2 − 3 % shear strain.

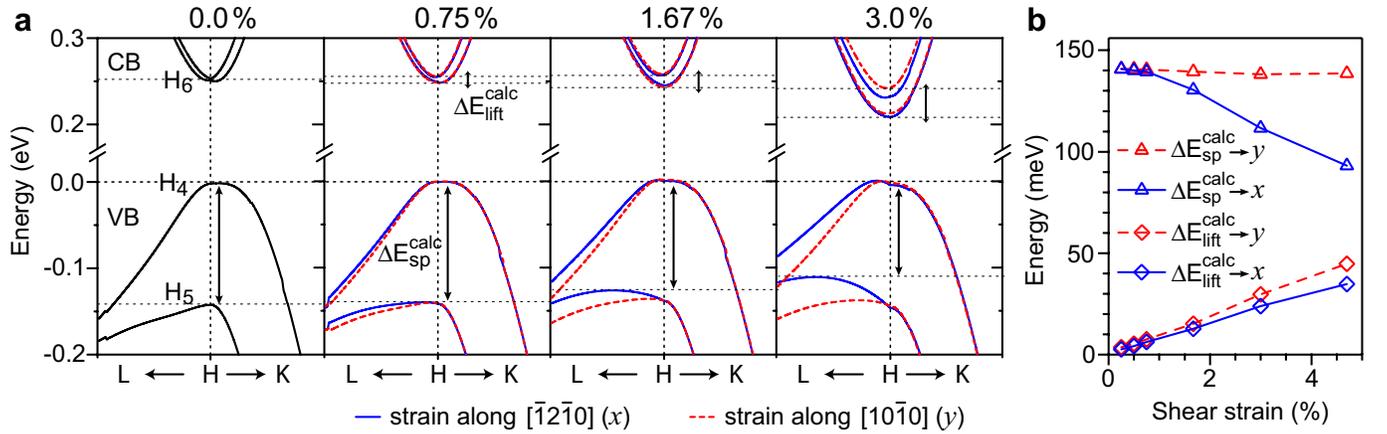

**Figure 4 | Te band structure under shear strain.** (**a**) Evolution of Te band structures around the H-point near the Fermi level with increasing the shear strain. Blue lines correspond to the shear strain along $[\bar{1}2\bar{1}0](x)$ while red dashed lines correspond to the strain along $[10\bar{1}0](y)$ directions, which is applied on the two (0001) planes of Te crystal. (**b**) Variations of spin degeneracy lifting $\Delta E_{lift}^{calc}$ of the $H_6$ CB (or $H_6^{CB}$) and the spin-split gap $\Delta E_{sp}^{calc}$ between frontier $H_4$ and $H_5$ VBs as a function of the magnitude of shear strain (written on the top in %) along both directions, as indicated by $x$ and $y$, respectively.

The photoinduced strain generation can be caused by thermoelastic, electron-acoustic deformation potential, or the inverse piezoelectric effect (IPE). Both the thermoelastic and deformation potential mechanisms produce small stresses ($< 10^{-3}$ GPa, see Supplementary Note 7) due to the small excitation density; too small to induce the tens of meV band-gap shifts[47] that we observe. We believe that the IPE is the most likely mechanism for shear strain generation. IPE has been observed in various piezoelectric semiconductors[49, 50] and is also expected in Te since it is an anisotropic crystal with moderately high piezoelectric coefficients[26, 27, 51] and predicted to show ferroelectricity.[52] Upon pump excitation, photogenerated carriers screen the built-in electric field in the film (most likely at the surface)[27] that induces a strain wave through IPE. The induced strain relaxes as excited carriers decay because the screening slowly diminishes. The photoinduced strain generation can be observed through a periodic change of the TR signal due to the interference between the propagating strain wave and the probing laser pulse.[53, 54] We have observed a weak oscillatory component with an average period of $26.3 \pm 0.1$ ps in our TR signal corresponding to the coherent longitudinal acoustic phonons (CLAP) propagating in the film, which supports photoinduced strain generation in our samples (see Supplementary Note 7). The sound velocity obtained is much lower than the longitudinal sound velocity along the c-axis or perpendicular to it, and rather closer to the velocity of the shear wave along the same directions. This may be an indication of shear strain contributions along with other modes in our samples.[55, 56]

**Polarization and energy-dependent carrier dynamics.** The transient polarized TR spectra enable study of multiple carrier decay processes. Polarized time decays from the low and high energy regimes are shown in Fig. 5a,b. Simple tri-exponential decay functions convoluted with a Gaussian response function (see Eq. 2 in Supplementary Note 3) fit the data very well over the entire delay range and provide a quantitative picture of overall decay behavior. Near the band-gap region (Fig. 5a), the initial fast and slow decay constants are $\tau_1^\perp \sim 5 \pm 1$ ps and $\tau_2^\perp \sim 20 \pm 1$ ps, respectively, for $E \perp c$ and both decay times are twice as long for $E \parallel c$, i.e., $\tau_1^\parallel \sim 9$ ps and $\tau_2^\parallel \sim 36$ ps, respectively. There is also a small (~ 1 %) residual signal that slowly decays within $\tau_3^\perp \simeq \tau_3^\parallel \sim 1$ ns. Around the high-energy regime (Fig. 5b), only a fraction (~ 30 %) of the peak intensity decays abruptly within a sub-ps ($\tau_1^\parallel \sim 0.8 \pm 0.1$ ps) time for $E \perp c$, and is nearly 10-times slower for $E \parallel c$. The majority of the signal at high energy decays slowly with a decay time on the order of ~ 1 ns for both polarizations.

The physical origin of the TR decays can be understood from the schematic band diagram in Fig. 5c. Femtoseconds after pump excitation (vertical red arrows), the TR response decreases abruptly due to absorption bleaching through state filling of hot carriers before a Fermi-Dirac distribution is established by carrier-carrier scattering. Subsequent recovery of the TR response occurs through basic carrier relaxation processes that vary at different bands due to the different scattering channels involved. The TR signal near the band-gap decays rapidly as hot carriers thermalize quickly with the lattice through a cascade of optical phonon emissions and carrier-carrier scattering, as indicated by curved arrows in Fig. 5c. The early transient is dominated by thermalization of hot holes due to efficient scattering with the cold degenerate hole gas due to their higher effective mass ($m_h > m_e$). The fast decay of $\tau_1^\perp \sim 5$ ps corresponds to an upper bound to the effective scattering time for a hole density of $N_d$. The subsequent slower decay $\tau_2^\perp \sim 20$ ps is attributed to the interband recombination near the band-gap. Since our samples are thinner than the laser penetration depth $l_{IR}$ in Te around the IR-region ($l_{IR} \sim 50$ nm),[28] carrier diffusion processes do not contribute to the decay response. The weak residual signal ($\sim 1$ %of the peak) persists over nanosecond times and is attributed to feeding of long-lived carriers from higher lying bands through phonon-assisted intervalley processes. The initial rapid decay for $E \parallel c$ is a factor of two slower than that for $E \perp c$, as shown more clearly in the insets of Fig. 5a. This observation is directly related to the anisotropic transport characteristics of Te. The electrical conductivity of bulk Te, including the hole mobility and scattering time, has been found to be larger ($\sim 1.2 - 2.3$ times larger) along the c-axis than that of perpendicular to the c-axis.[15, 57] In particular, the free carrier scattering time parallel to the c-axis is measured to be 3-times longer than that perpendicular to the c-axis due to stronger coupling of polar optical scattering perpendicular to the c-axis, which is the main scattering process in *p*-doped Te samples.[15, 57] Such an anisotropic carrier scattering process is nicely reflected in the carrier dynamics of our samples, which could be also relevant in other anisotropic materials.

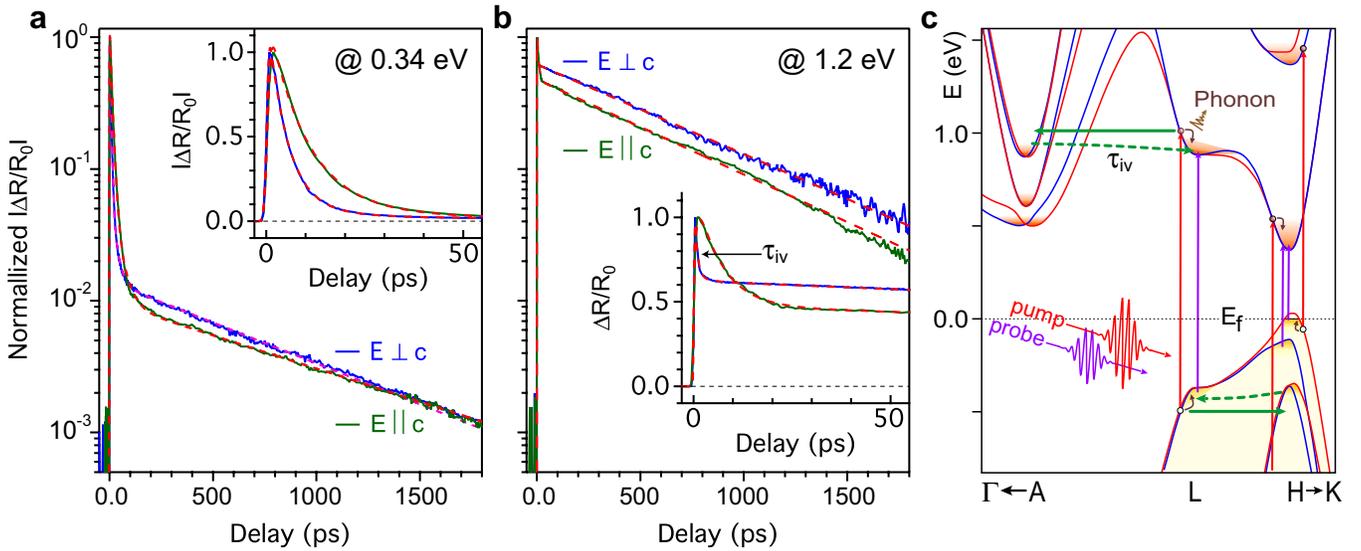

**Figure 5 | Polarization and energy dependent carrier decay dynamics.** Representative polarization-resolved TR ($\Delta R/R_0$) traces of Te samples around the fundamental gap (**a**) and higher energy transition region (**b**). Dashed red lines are the multi-exponential fits of each corresponding data over a long delay range, as described in the main text. Insets show the initial decay responses of each corresponding trace. Around band-edge region, a majority of the TR signal relaxes within first 30 ps due to ultrafast intraband thermalization followed by interband recombination. Around the higher energy transition, only a fraction of the maximum TR signal decays rapidly predominantly by intervalley scattering accompanied by intraband thermalization, while the rest of the signal recovers on the order of a nanosecond due to interband recombination. (**c**) Schematic drawing of carrier excitation and relaxation processes within extended bulk band structure of Te around high symmetry points in the Brillouin zone. Pump excitation is indicated by vertical red arrows and probe excitation is shown by vertical purple arrows. Intervalley forward and return scattering processes are also shown by green solid and dashed arrows.

The carrier decay dynamics at high energies is associated with the direct transition near the band-edge at the L-point and, therefore involves different scattering processes. Due to closely spaced CBs at the A-valley (see Fig. 5c), intervalley coupling of hot carriers between L- and A-valley is highly favorable. Photoexcited hot carriers at the L-valley will scatter rapidly to the A-valley via deformation potential scattering followed by return scattering to the L-valley, as indicated by green solid and dotted arrows, respectively, in Fig. 5c. The TR response for $E \perp c$ displays sharp transient within a sub-ps time ($\tau_1^\perp \sim 0.8$ ps) followed by slower recovery of the signal ($\tau_2^\perp \sim 5$ ps) until it reaches thermal equilibrium with the lattice (quasi-steady state regime). We attribute the early sub-ps decay time to an upper limit of average intervalley scattering time $\tau_{iv}$ between the L- and A-valleys and subsequent slower decay constant to the intraband cooling time within the L-valley, which is nearly identical with the band-edge thermalization time at the H-point. Our data of $\tau_{iv} \sim 0.8$ ps is comparable with the data obtained from other semiconductors such as GaAs.[58] Since the intervalley return scattering is much longer than the forward scattering due to the lower density of states at the A-valley, the second decay component $\tau_2^\perp$ may involve both intervalley return scattering as well as intraband cooling processes. The TR signal after cooling of carriers, i.e., the degree of state filling of thermalized carriers at the band-edge, reduces only marginally (only ~ 30 % of the peak), which is likely due to the reduced density of optically coupled state filling at the band minima due to dominant intervalley forward scattering. The ultrafast intervalley process creates a long-lived reservoir of carriers at the A-valley, which eventually bleeds back to the band minima at the H-valley. This extraction time corresponds to the long-lived decay ($\tau_3^\perp \sim 1$ ns) seen for the H-valley transition (see Supplementary Note 4.) The TR response for $E \parallel c$ qualitatively resembles the subsequent long-lived decay response (see Fig. 5b). The early decay is dominated by the much slower ($\tau_1^\parallel \sim 8$ ps) carrier thermalization rate, which is again, as discussed earlier, supported by the arguments of higher mobility and scattering lifetime of free carriers along the c-axis. This observation of longer early decay response suggests intervalley scattering is effectively suppressed with $E \parallel c$ and hints also at possible consequences of photoinduced band modification even in the higher lying bands.

**Discussion**

Optical transitions in Te are found to be highly anisotropic and strongly perturbed upon photoexcitation, which are linked to its unique spin texture and band dispersion. The detailed analysis of transient reflectivity spectra at late times reveals direct degenerate transitions at the band-gap between the Fermi level, 30 meV below the top of the $H_4$ band-edge, and the $H_6^{CB}$ for both polarizations and direct transition between $H_5 \to H_6^{CB}$ only for $E \parallel c$, which agree well with bulk absorption measurements.[15]

While such a result has been seen by others,[15, 31] it is inconsistent with the group theory result that the $E \parallel c$ transition should be forbidden at the fundamental gap.[28] This disagreement may reflect the unique spin texture and complex band dispersion near the band-edge. The data also shows an indirect transition between the doubly degenerate $H_6^{VB} \rightarrow H_6^{CB}$ at ~ 0.7 eV, which has been predicted theoretically but not previously observed experimentally. In addition, we also identify a possibly direct transition at ~1.2 eV at the L-point band-edge. Detailed energy-dependent time traces show ~99% of the photoexcited carrier decay through recombination at the band-gap within 30 ps while ~1% of the photoexcited carriers are scattered from the VB or CB at the L-point to a remote valley likely at the A-point in the Brillouin zone. These trapped carriers gradually feed from the remote valley over the following 1 ns back to the CB and VB band-edges at the H-point where they rapidly recombine. These highest energy transitions appear to be long-lived suggesting they can be a source for charged carriers, potentially important in the design of efficient thermoelectric and opto-electronic devices.

The detailed analysis of TR spectral dynamics reveals ~2% photoinduced shear strain through IPE, which breaks the screw symmetry of the Te and lifts the degeneracy of the $H_6^{CB}$ (~ 20 meV splitting). This causes also ~ 30 meV decrease in the spin-split gap between the $H_4$ and $H_5$ VBs due to strain induced reduction in the SOI. The bands relax to equilibrium as the photoexcited carriers relax within 30 ps. These results are confirmed by detailed ab initio DFT calculations as a function for strain, which indicate that the photoinduced shear strain is ~ 2 %. While such a strain is too low to achieve a topological phase transition, [8, 24] these results provide solid evidence that it is possible to manipulate symmetry protected energy states through photoexcitation. We find that even relatively modest photoinduced shear strains in Te can lift the degenerate crossing of the Weyl point at the $H_6^{CB}$. Future research into the response of quantum materials (e.g., Weyl semimetals) to intense illumination may be quite promising.

## Methods

**Growth of Te nanosheets.** Te nanosheet samples were grown chemically through the reduction of sodium tellurite ($Na_2TeO_3$) by hydrazine hydrate ($N_2H_4$) in an alkaline solution at temperatures 160 – 200°C with the presence of crystal-face-blocking ligand polyvinylpyrrolidone (PVD).[12] The hydrophilic Te nanosheets were transferred to thermally oxidized Si substrates by the Langmuir–Blodgett assembly process.

**Sample characterization**. The surface morphology and thickness of Te sheets were measured by using atomic force microscope (AFM). Measurements were conducted in air on a Veeco AFM using gold (Au) coated platinum (Pt) cantilevers operating in tapping mode. Raman spectroscopy was applied to characterize sample quality and crystal orientation. The measurements were performed using a Raman microscope with 632.8 nm laser excitation in the backscattering configuration under ambient condition. The Te sample was rotated with respect to the laser polarizations in order to identify the crystal c-axis and characterize polarization dependent Raman spectra. A 100× objective was used to focus the incident beam and collect the scattered signal and then dispersed by 1800 g/mm grating after the resonant scattered laser light is removed using a double subtractive mode spectrometer. The Raman signal is measured with a spectral resolution of 1 cm$^{-1}$. Laser power was maintained around 100 μW during the measurements to minimize laser-heating effect in our samples.

**Transient reflectance spectroscopy**. Ultrafast pump-probe TR measurements were performed based on a Ti:sapphire oscillator which produces 150 fs pump pulses at a central wavelength of 800 nm with 80 MHz repetition rate and an average power of 4 W. The majority (80%) of the laser beam was used to pump an optical parametric oscillator (OPO), which generates signal (0.72 − 1.2 eV) and idler (0.3 − 0.8 eV) photons, which can be tuned continuously over a wide range. All measurements were performed using 820 nm ($\hbar\omega_p = 1.51$ eV) as the pump pulse and both signal and idler outputs as probe pulses, respectively, as illustrated in the Supplementary Figure 3. The polarization of the probe pulses with respect to the crystal c-axis is controlled using a CaF$_2$ double Fresnel rhomb on the probe beam path. The probe pulses are delayed in time with respect to the pump pulses by using a motorized linear translation stage. The pump and probe beams are spatially overlapped on a single Te sheet using a 40× protected silver reflective objective with the help of a CCD camera, and the reflected beams are directed to the liquid nitrogen cooled InSb detector. The pump beam is filtered out during the TR measurements using a long pass filter. The pump-induced signal is collected with a lock-in amplifier phase-locked to an optical chopper that modulates the pump beam at a frequency of 1 kHz. The focused spot size of both the pump and probe beams is nearly identical, i.e., ∼ 2 μm in diameter. An average pump fluence of ∼ 0.5 − 1 mJ/cm$^2$ was used throughout the experiments, while the probe fluence was kept nearly one order of magnitude lower.

**Strain-induced band structure calculations:** The calculation of the Te band structure under strain is performed based on density functional theory (DFT) as implemented in the Vienna ab initio Simulation Package (VASP).[59] The projector-augmented wave method[60, 61] is employed for DFT calculations, with the Perdew–Burke–Ernzerhof (PBE)[62] generalized gradient approximation (GGA)[63] for the exchange–correlation functional. The spin-orbit coupling is considered in all DFT calculations. The Tkatchenko and Scheffler (DFT-TS) method[64] was employed for the van der Waals correction which provides excellent agreement with experimental lattice constants of the bulk Te.[65, 66] Other van der Waals correction methods such as DFT-D2[67] and DFT-D3[68] were also used to check against the DFT-TS results but negligible differences are found, therefore, not reported in this paper. The Γ-centered Monkhorst-pack[69] k-point sampling is used for the Brillouin zone integration. We find that the 10×10×8 grid is sufficient for the convergence. The plane-wave cutoff energy is set to 500 eV and the convergence criterion for electronic relaxations is set to $10^{-6}$ eV. All atoms in the computational cell are relaxed until the Hellmann–Feynman forces are less than 0.01 eV Å$^{-1}$. The band structures were obtained using the HSE06 functional.[70, 71]

## Data availability

The data that support the plots within this paper and other findings of this study are available from the corresponding authors upon reasonable request.

## Acknowledgments

HEJ and LMS acknowledge the financial support of the NSF through grants DMR 1507844, DMR 1531373, ECCS 1509706, while YSL acknowledges the NSF through grant ECCS 1609902. CL and BMW acknowledge support by the DOE through grant DE-SC0016269. P.D.Y. acknowledges NSF/AFOSR 2DARE, SRC and DARPA for their support. W.Z.W. and P.D.Y. are also supported by NSF through grant CMMI 1762698 and ARO through grant W911NF-17-1-0573. W.Z.W. acknowledges the Ravi and Eleanor Talwar Rising Star Assistant Professorship from the School of Industrial Engineering at Purdue University.

## Author contributions

G.J. and L.M.S. conceived the study. R.W., G.Q, W.Z.W., and P.D.Y. synthesized Te nanosheet samples. G.J. characterized the samples, performed TRS measurements, data analysis/modeling, figure planning, and draft preparation. Y. X. and Y.S. L. performed ab initio DFT calculations as a function of uniaxial and shear strain. C.L. and B.M.W. performed preliminary RT-TDDFT calculations. I.A.S. helped in the micro-Raman measurements. S.L. helped in the TRS experiments and modeled the TRS spectra with inputs from G.J., L.M.S., and H.E.J. G.J. and L.M.S. wrote the manuscript. All coauthors contributed to the discussion of results and commented on the manuscript.

## Corresponding authors

Correspondence to Giriraj Jnawali (E-mail: giriraj.jnawali@uc.edu) or Leigh M. Smith (E-mail: leigh.smith@uc.edu).

## Competing interests

The authors declare no competing interests.

# Supplementary Information

**Ultrafast Photoinduced Band Splitting and Carrier Dynamics in Chiral Tellurium Nanosheets**

Giriraj Jnawali *[1], Yuan Xiang [2], Samuel M. Linser [1], Iraj Abbasian Shojaei [1], Ruoxing Wang [3], Gang Qiu [4], Chao Lian [5], Bryan M. Wong [5], Wu Wenzhuo [3], Peide D. Ye [4], Yongsheng Leng [2], Howard E. Jackson [1] and Leigh M. Smith*[1]

[1] *Department of Physics and Astronomy, University of Cincinnati, Cincinnati, OH 45221, USA*
[2] *Department of Mechanical & Aerospace Engineering, The George Washington University, Washington, D.C. 20052, USA*
[3] *School of Industrial Engineering, Purdue University, West Lafayette, IN 47907, USA*
[4] *School of Electrical and Computer Engineering, Purdue University, West Lafayette, IN 47907, USA*
[5] *Department of Chemical & Environmental Engineering, Materials Science & Engineering Program, University of California, Riverside, Riverside, CA 92521, USA*

*Corresponding authors

## Supplementary Note 1: Strain effects on the electronic band structure of Te

### A. Band structure under zero strain

The Te band structure is calculated using density functional theory (DFT) as implemented in the Vienna ab initio Simulation Package (VASP).[1] Computational details are described in Methods section of the main text. Supplementary Figure 1 shows the calculated Te band structure around high symmetry points in the Brillouin zone. All the features in the band structure are reproduced by our calculations (see main text for details). In particular, the computed band gap is at the H-point with the magnitude of 0.25 eV, which is somewhat lower than the experimental value of 0.32 eV but is in good agreement with other DFT results.[2, 3] The two-fold spin degeneracy of the $H_6$ conduction band (CB) or conventionally $H_6^{CB}$ due to the three-fold screw symmetry of Te is clearly seen. There are two non-degenerate (spin-polarized) $H_4$ and $H_5$ valence bands (VBs) with splitting gap of about 0.14 eV, which is consistent with our observation of the energy-splitting between the $H_4$ and $H_5$ band edges, i.e., $\Delta E = 0.115$ eV, as well as recent angle-resolved photoemission spectroscopy (ARPES) mapping of VBs of single crystal Te.[4] Not only the low-energy bands but also higher lying bands such as $H_6$ VB or $H_6^{VB}$ as well as L- and A-valley structures are also consistent with previous DFT studies, which are explicitly discussed in the main text. Next, we discuss about how these features, in particular near the Fermi energy around the H-point, are affected by the strain applied in the Te crystal along different crystallographic directions.

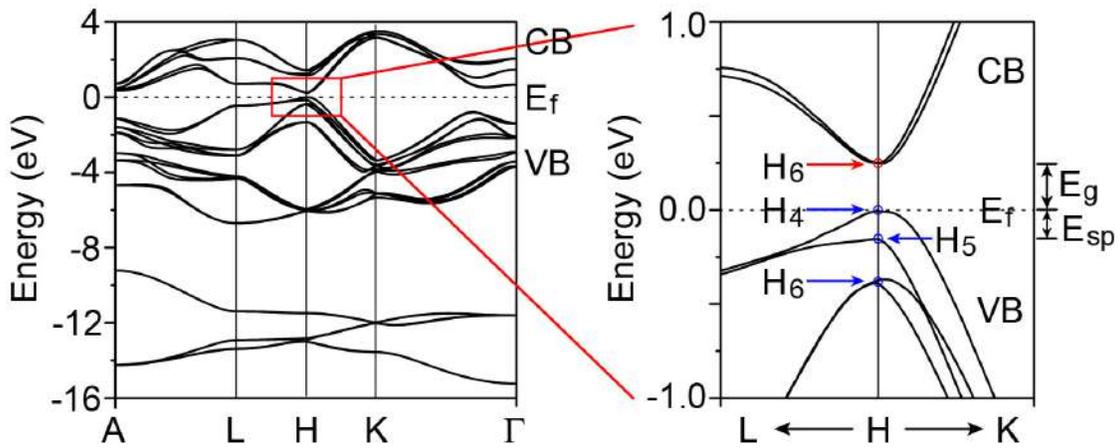

**Supplementary Figure 1 | Computed band structure of Te under zero strain.** The band structure is calculated along the symmetry directions of the Brillouin zone of the bulk Te. The Band structure is particularly relevant around the H-point near the Fermi level $E_f$ where the band-gap $E_g$ exists, as shown by the enlarged diagram on the right. The entire band features, including the two-fold spin degeneracy at $H_6$ CB, the spin-polarized non-degenerate frontier $H_4$ and $H_5$ VBs (with spin-split gap $E_{sp}$), the two-fold spin degeneracy of $H_6$ VB, as well as the slightly indirect band-gap between $H_6$ CB and $H_6$ VB along the H-K direction, are consistent with previous works.[2, 3, 5, 6]

## B. Band structure under uniaxial strains or a hydrostatic pressure applied

It is well known that the band structure of a solid crystal is modified by inducing strains. The band modification, however, depends on how the crystal symmetry is affected by different types of strains. In the case of Te, an anisotropic chiral semiconductor, its complex band structure near the band-edge displays unique strain effects that have been demonstrated by recent studies of ab initio electronic structure calculations.[2, 3, 6] Applying uniaxial strains or a hydrostatic pressure on a Te crystal (see Supplementary Figure 2a), we see that the band-gap near the H-point in the Brillouin zone increases in the case of a compressive strain applied and decreases in the case of a tensile strain or a hydrostatic pressure applied. Very large tensile strains or hydrostatic pressures can lead to the closing of the band-gap or even a band crossing of the frontier bands (band inversion). However, the three-fold symmetry of the helical structure is preserved and thus the two-fold degeneracy of the $H_6^{CB}$ band is maintained. These results are shown in Supplementary Figure 2b.

When a shear strain is applied, however, the screw symmetry of the helical chains is broken, which lifts the degeneracy of $H_6^{CB}$, as well as reduces the spin-split gap of the uppermost VBs in the vicinity of H-point. We have discussed this scenario in the main text.

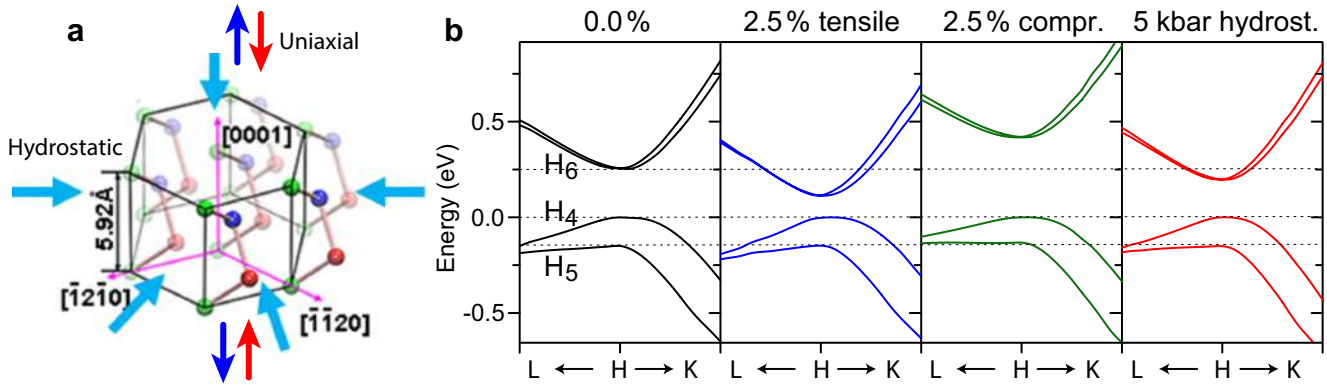

**Supplementary Figure 2 | Computed band structures of Te under uniaxial strains or a hydrostatic pressure. a** Schematic of various strains applied along different directions of the Te crystal. The dark blue and red arrows show the tensile and compressive uniaxial strains applied along the c-axis, respectively. The light blue arrows show a hydrostatic pressure applied on the crystal from all directions. **b** Changes of the band structure around the H-point near the Fermi level without and with various strains (plotted in different colors). The types and magnitudes of the strains are also shown on the top of each plot.

## Supplementary Note 2: Polarized transient reflectance spectroscopy setup

The standard pump-probe setup used for polarized transient reflectance spectroscopy measurements is shown schematically in Supplementary Figure 3. The Si/SiO$_2$ substrate with the Te nanosheet is mounted to the 10 K cold finger of a microscope-cryostat. A short 1.51 eV pump pulse (150 fs) is superimposed collinearly with a probe pulse (150 fs) of variable photon energies (0.3 − 1.2 eV) and focused onto a single Te nanosheet using a 40 × 0.5 NA reflective objective. The electric field polarization ($\phi$) of the incident probe beam is controlled using a CaF$_2$ double Fresnel Rhomb rotator on the probe beam path. Measurements are performed with probe polarization either parallel ($E \parallel c$), i.e., $\phi = 0°$ or perpendicular ($E \perp c$), i.e., $\phi = 90°$ with respect to the c-axis of the Te crystal. The crystal c-axis is defined based on polarization dependent Raman measurements on a single Te nanosheet samples. For steady-state reflectance measurements, the probe reflectance is detected with pump off conditions. For pump-induced transient reflectance measurements the probe reflectance is detected with pump on conditions in which the probe pulses are delayed in time with respect to the pump pulses by using a motorized linear translation stage. The reflectance signal is detected using the liquid nitrogen cooled InSb detector. The probe pulses are delayed in time with respect to the pump pulses by using a motorized linear translation stage. The pump beam is filtered out during the TR measurements using a long pass filter. The pump-induced signal is collected with a lock-in amplifier phase-locked to an optical chopper that modulates the pump beam at a frequency of 1 kHz.

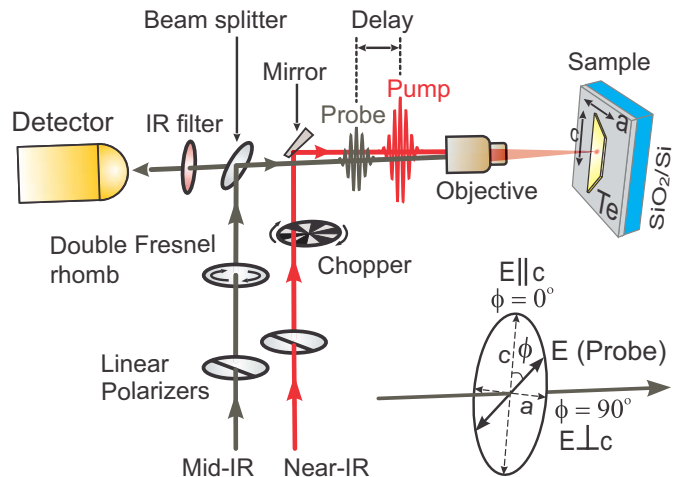

**Supplementary Figure 3 | Polarized transient reflectance spectroscopy (TRS) setup.** Simple illustration of polarized TRS experiment performed in Te nanosheets, which are deposited on SiO$_2$/Si substrate. The polarization state of the probe beam with respect to the crystal c-axis is controlled by using a double Fresnel rhomb. Time and probe energy dependent pump-induced reflectance spectra are measured at each orthogonal polarizations, as shown on the bottom right, using standard pump-probe detection technique.

# Supplementary Note 3: Transient reflectance spectroscopy (TRS) at 300 K

## A. Optical transitions at 300 K

Supplementary Figure 4a displays a false color map of the 300 K TR spectra measured over a wide spectral region (0.3 − 1.2 eV) for the probe polarized parallel ($E \parallel c$) and perpendicular ($E \perp c$) to the c-axis. Overall, the TRS map exhibits strong anti-symmetric features around low-energy (0.3 − 0.5 eV) and high-energy (0.8 − 1.2 eV) regions and a relatively weaker and broader signal near the mid-energy (0.5 − 0.8 eV) region, corresponding to a series of interband optical transitions. The map also shows that the strong features remain essentially unchanged in shape with time except for their magnitudes, suggesting only a weak perturbation of the transition response by the carrier thermalization process. These features are more apparent in Supplementary Figure 4b where the spectral slices at 50 ps delay from each polarization are plotted (blue circles). The derivative-like spectral features of the transient signal around the low- and high-energy regions exhibit significant differences for the two polarizations, while the broad feature in the mid-energies also displays some sensitivity to polarizations. With $E \parallel c$, a nearly overlapped additional feature is observed at ∼0.44 eV, which is not present with $E \perp c$, suggesting the polarization sensitive optical absorption in the Te samples.

It is known that the TR response is related to the perturbation of the dielectric function of the material induced by photoexcitation of charged carriers. For the Lorentzian form of the dielectric function under low-field modulation with the parabolic band approximation, TR spectra around the band-edge regions can be analyzed by using a derivative Lorentzian lineshape functional form appropriate for excitonic transitions,[7]

$$\frac{\Delta R}{R_0}(E) \simeq \sum_{j=1}^{n} Re\left[A_j e^{i\varphi_j}(E - E_j + i\Gamma_j)^{-2}\right], \qquad (1)$$

where $n$ represents the number of spectral functions for the possible interband transitions involved, $E$ is the probe photon energy, and $A_j$, $\varphi_j$, $E_j$, and $\Gamma_j$ are the amplitude, phase, transition energy, and the energy broadening parameter of the $j^{th}$ feature, respectively. In order to quantify each of these features and corresponding transition energies, the TRS spectra are fitted using Eq. 1, as shown with red dashed-lines in Supplementary Figure 4b. The spectra with $E \perp c$ can be fitted by three resonances ($n = 3$) while the spectra with $E \parallel c$ can be fitted by four resonances ($n = 4$). Moduli of the individual resonances, which have been obtained from Eq. 1, are also overlapped with minor vertical shifts for clarity (black solid lines in Supplementary Figure 4b). The values of the transition energies obtained from least square fitting are indicated by vertical dashed grey lines.

Now, we compare the transition energies of each $j^{th}$ feature with the detailed band structure around the H–point of the bulk Brillouin zone of Te,[5, 6, 8] as shown schematically in the main text. The lowest energy transition $H_4 \to H_6^{CB}$ with $E \perp c$ is $E_1^\perp = 0.34 \pm 0.002$ eV, which is nearly the same value as the low temperature value shown in the main text. This behavior nicely corresponds with the peculiar temperature dependent absorption coefficient in Te previously observed.[9] One can also see the lowest energy transition with $E \parallel c$, which is blue-shifted by $\sim 20$ meV to $E_1^\parallel = 0.36 \pm 0.002$ eV. Such a polarization anisotropy of optical transition at the band-gap remains unchanged over entire delays, in contrast to the strong modulation up to 30 ps observed at 10 K (see main text). The $E_1^\perp$ transition energy as well as the polarization anisotropy, i.e., the dichroism of the optical absorption edge, agree very well with previous results obtained by linear absorption measurements on degenerately p-doped Te samples at room temperature.[9-11] The optical transition between $H_5 \to H_6^{CB}$ is observable only with $E \parallel c$ and totally absent with $E \perp c$, which is in agreement with low temperature measurements and consistent with the expected dipole allowed transition between these two states. The $H_5 \to H_6^{CB}$ transition energy is estimated to be $E_2^\parallel = 0.44 \pm 0.005$ eV, which is again nearly identical with the value measured at 10 K. The spin-split VB gap or the separation between the $H_4$ and the $H_5$ VBs turns out to be $E_2^\parallel - E_1^\perp = 100$ meV. This value, taking into account the doping-induced Fermi level shift, is nearly the same as the so-called 11-micron hole absorption band previously measured in a bulk single crystal of Te.[12, 13] The energy difference between the $E_2^\parallel$ transition and $E_1^\parallel$ transition decreases to $E_2^\parallel - E_1^\parallel = 80$ meV, which suggests not only the fundamental gap but also the gap between spin-split VBs is anisotropic at room temperature. In addition to the direct transitions between uppermost VBs and the lowest CB, there is an indication of a higher energy transitions at around $E_3^{\parallel,\perp} = 0.75 \pm 0.005$ eV with the lineshape sensitive to polarizations. Since the TR signal is nearly one order of magnitude weaker than the near band-edge transition, this transition might be indirect in **k**-space, as expected from the band structure.

We also observed a higher lying transition at $E_4^\perp \sim 1.16 \pm 0.005$ eV, which is blue shifted by 20 meV to $E_4^\parallel \sim 1.18 \pm 0.005$ eV. Due to limitations of the spectral range of the probe pulse, a full derivative-like lineshape is not observed, which causes uncertainty in the precise energy transition. Nevertheless, the transition is clearly sensitive to polarizations and the anisotropy is the same as in the fundamental gap. This high-energy transition is noticeably red-shifted (by $\sim 80$ meV for $E \perp c$ and $\sim 60$ meV for $E \parallel c$) as compared to the value observed at low temperature $E_4^{\parallel,\perp} \sim 1.24 \pm 0.005$ eV, suggesting fundamentally different in nature. Since the CBs above $H_6^{CB}$ and VBs below $H_6^{VB}$ at the H-point are far apart, the next higher energy transition at the H-point beyond $E_3^{\parallel,\perp}$ is not accessible with our probe beam. Therefore, we attribute the transitions $E_4^\perp$ and $E_4^\parallel$ are caused by a direct transition at the L-point. Detail understanding of higher energy transitions in Te requires further theoretical calculations.

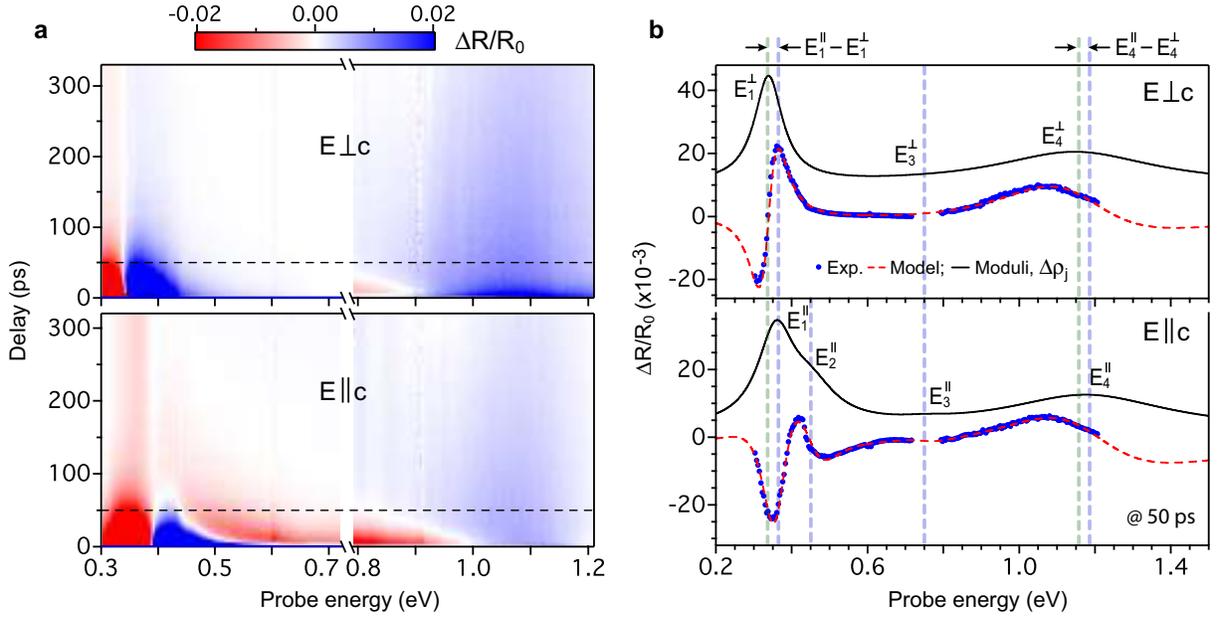

**Supplementary Figure 4 | Polarized transient reflectance response of Te nanosheets around the band-edge region. a** Two-dimensional false color map of pump-probe delay dependent transient reflectance spectra $[\Delta R/R_0 (E, t)]$ over an extended probe energies for parallel $E \parallel c$ and perpendicular $E \perp c$ relative to the c-axis of Te crystal. Narrow region of probe energies between $0.72 - 0.8$ eV is missing due to technical limitations. Negative signal (red) indicates a pump-induced increase in absorption and positive signal (blue) corresponds to decrease in absorption. Overall, strong features with distinctive dynamics at different energies are clearly visible, suggesting a series of optical transition with some polarization anisotropy. **b** Transient spectra (blue spheres) acquired at 50 ps time delay with two orthogonal probe polarizations. Each spectrum is fitted with a simple model described in the main text (red dashed lines). Calculated moduli $\Delta \rho_j$ of each fit are also plotted with vertical offset (black lines). Both the model fits and the moduli show transient features corresponding to the series of direct and indirect interband transitions in Te, as indicated by vertical dashed lines. The transition energy for $E \parallel c$ is blue-shifted by $\sim 20$ meV, suggesting band-edge optical anisotropy in Te at room temperature.

## B. Carrier dynamics at 300 K

Polarization dependent carrier decay processes are investigated by measuring TR time traces at different probe energies following excitation of the sample with 1.51 eV pump pulses. Representative polarized time traces taken from the nanosheet at 300 K from two low- and high-energy regimes are shown in Supplementary Figure 5a,b. At energies intermediate between these two regimes, the TR response is very weak and fast due to lack of any direct optical transitions and, therefore we will not discuss it here. Near the band gap region (Supplementary Figure 5a), the majority of the TR decays exponentially within first 30 ps followed by a very weak (nearly 2-orders of magnitude lower than the peak) residual signal, which persists over 300 ps. Around the high-energy regime (Supplementary Figure 5b), the time traces display initial ultrafast decay of the signal followed by a long exponential recovery of the remaining signal at later times. In order to quantify the overall decay behavior around the low and high energy regimes, each time trace is fitted using multi-exponential functions convoluted with a Gaussian response function:[14]

$$\Delta R/R_0 (t) = \sum_{i=1,2,3} c_i \cdot \sigma \cdot exp[(\sigma/2\tau_i)^2 - (t/\tau_i)] \cdot [1 - erf\{(\sigma/2\tau_i) - (t/\sigma)\}], \quad (2)$$

where $c_i$ is the amplitude with decay time constant of $\tau_i$ of the $i^{th}$ exponential term and $\sigma$ is full width half maximum of the pump laser pulse ($\sigma = 200\ fs$). A least square fitting to Eq. (2), as shown by dashed red lines in Supplementary Figure 5a,b estimates the decay time constants $\tau_i$ of each $i^{th}$ decay channel at different bands. Near the low-energy region, the decay constant of the majority of the signal is $\tau_1^\perp \sim 17 \pm 1$ ps for $E \perp c$ and $\tau_1^\parallel \sim 22 \pm 1$ ps for $E \parallel c$, respectively, followed by long-lived ($\tau_2^\perp, \tau_2^\parallel \gtrsim 500$ ps) residual signal. Around the high-energy region, only a fraction of the peak intensity decays abruptly within a few ps ($\tau_1^\perp \sim 10 \pm 1$ ps for $E \perp c$ and $\tau_1^\perp \sim 12 \pm 1$ ps for $E \parallel c$) and the remaining signal decays rather slowly ($\tau_2^\perp, \tau_2^\parallel \gtrsim 300$ ps) for both polarizations. The fractional signal of rapid decay is $\sim 50\ \%$ of the peak signal for $E \perp c$ while it is only $\sim 10\ \%$ of the peak signal for $E \parallel c$. Such differences of decay dynamics with respect to the polarization and energy of the probe laser beam are qualitatively similar as observed at low temperature (see main text).

Around the low-energy region, ultrafast carrier thermalization by carrier-carrier and carrier-phonon scattering is not distinguishable at room temperature. Therefore, the decay time constants of $\tau_1^\perp$ and $\tau_1^\parallel$ for both polarizations are attributed to the interband carrier recombination time including carrier thermalization. Due to ultrathin samples, thinner than the penetration depth $l_{IR}$ in Te around the IR-region ($l_{IR} \sim 50$ nm),[9] carrier diffusion does not play a role in the decay dynamics. The subsequent weak residual signal (less than 2 % of the peak) observed after the recombination is attributed to a constant feeding of carriers from higher lying bands through phonon-assisted intervalley processes. Around the high-energy region, the polarized TR responses display sharp transient within $\sim 10$ ps, which is attributed to intervalley scattering followed by the intraband cooling of carriers via carrier-carrier and carrier-phonon scattering. Since the signal after initial rapid decay is reduced substantially with $E \perp c$ as compared to the signal with $E \parallel c$, this observation suggests intervalley scattering is effectively suppressed with $E \parallel c$. Subsequent slower decay of the signal is caused by persistent feeding of intervalley scattered carriers at neighboring valleys as well as carriers at higher lying bands to the H-valley. Note that the initial rapid recombination time at all spectral regions is noticeably slower with $E \parallel c$ as compared to the times with $E \perp c$. This behavior is more apparent at low temperature (see main text), which is consistent with anisotropic carrier scattering times and hole mobility observed in Te crystal.[10] Overall, carrier decay dynamics at room temperature is qualitatively similar to the low-temperature dynamics, which has been described with sufficient detail in the main text.

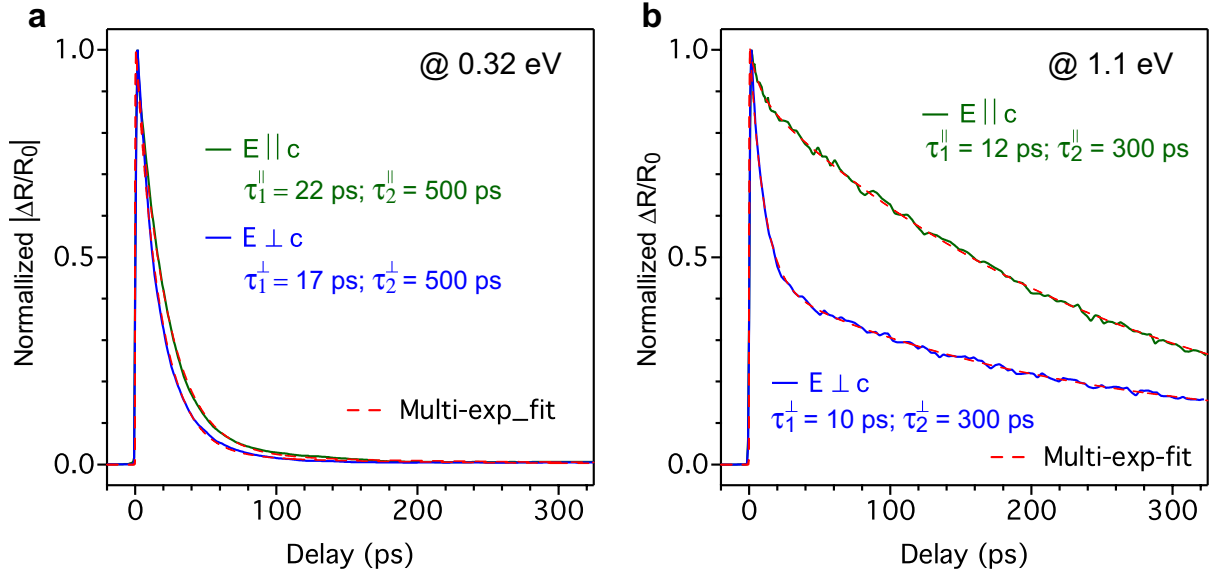

**Supplementary Figure 5 | Polarized transient reflectance response of Te nanosheets at room temperature.** Representative polarization resolved transient reflectance (TR) $\Delta R/R_0$ traces of Te samples around the fundamental band-edge **a** and high-energy transition region **b**. Dashed red lines are the multi-exponential fits of each corresponding data over a long delay range. Around the band-edge region, majority of the TR signal relaxes within first 30 ps due to ultrafast intraband thermalization followed by interband recombination. Around the higher energy transition, about half of the peak TR signal decays rapidly ($\tau_1^\perp \sim 10$ ps) with $E \perp c$ due to intervalley scattering accompanied by intraband thermalization. In contrast, TR signal with $E \parallel c$ decays only marginally (10 % of the peak) with a decay constant of $\tau_1^\parallel \sim 12$ ps, suggesting suppressed intervalley scattering and weak coupling to phonon. Rest of the signal for both polarizations decay rather slowly ($\tau_2^\perp, \tau_2^\parallel \gtrsim 300$ ps) due to constant feeding of carriers from higher lying bands to the H-valley.

## Supplementary Note 4: Modeling carrier recombination dynamics at 10 K

One of the major results described in the paper is that a certain fraction of carriers at high energy scatter into remote valleys away from the H-valley minimum. These long-lived carriers then can provide a long-lived source for carriers which eventually recombine at the band edge. In order to make a more quantitative analysis of the decay dynamics in this system we perform modeling of time decays using coupled rate equations:

$$\frac{d}{dx}n_h(t) = G_0 - \frac{n_h(t)}{\tau_{h0}} - \frac{n_h(t)}{\tau_h}; \quad \frac{d}{dx}n_l(t) = G_1 - \frac{n_l(t)}{\tau_l} + \frac{n_h(t)}{\tau_h}, \quad (3)$$

where $n_h(t)$ and $n_l(t)$ denote the number density of thermalized carriers at the high (presumably the indirect L-valley) and low-energy (the lowest energy H-valley) regimes. Considering a two-level system, the first equation determines the scattering time $\tau_h$ of which describes the slow feeding of carriers from neighboring higher lying valleys to the H-valley as well as lifetime $\tau_{h0}$ of residual carriers in those higher lying indirect valleys,, which is beyond the experimental limit and assumed to be $\tau_{h0} = 5$ ns. In other words, we assume the lifetime of the indirect higher lying valleys is determined by feeding carriers to the H-valley band edge. The second equation describes the direct recombination time $\tau_l$ with accounting for constant feeding of carriers from higher energy bands to the band edge minima. The terms $G_0$ and $G_1$ determine the fraction of carriers excited into indirect higher lying valleys vs. the fraction excited into the H-valley minima. The solution of Eq. (3) fits the experimental time traces around both regions reasonably well, as shown by dashed red lines in Supplementary Figure 6a,b. We determine that the ratio $G_0/G_1 \sim 1.6\%$ as expected from the time decays. The fits, however, does not cover the early transients (below 10 ps), where the carrier relaxation mostly dominated by carrier thermalization processes. The extracted values of carrier recombination times $\tau_l$ are , and ~15 ps and ~25 ps for $E \perp c$ and $E \parallel c$, respectively, and the lifetime of slow bleeding of carriers from higher lying valleys to H-valley for both polarizations is $\tau_h \approx 750$ ps. These values are nearly identical with the simple exponential fittings (shown in Supplementary Note 3) and agree with the previously described qualitative understanding of carrier decay dynamics in Te.

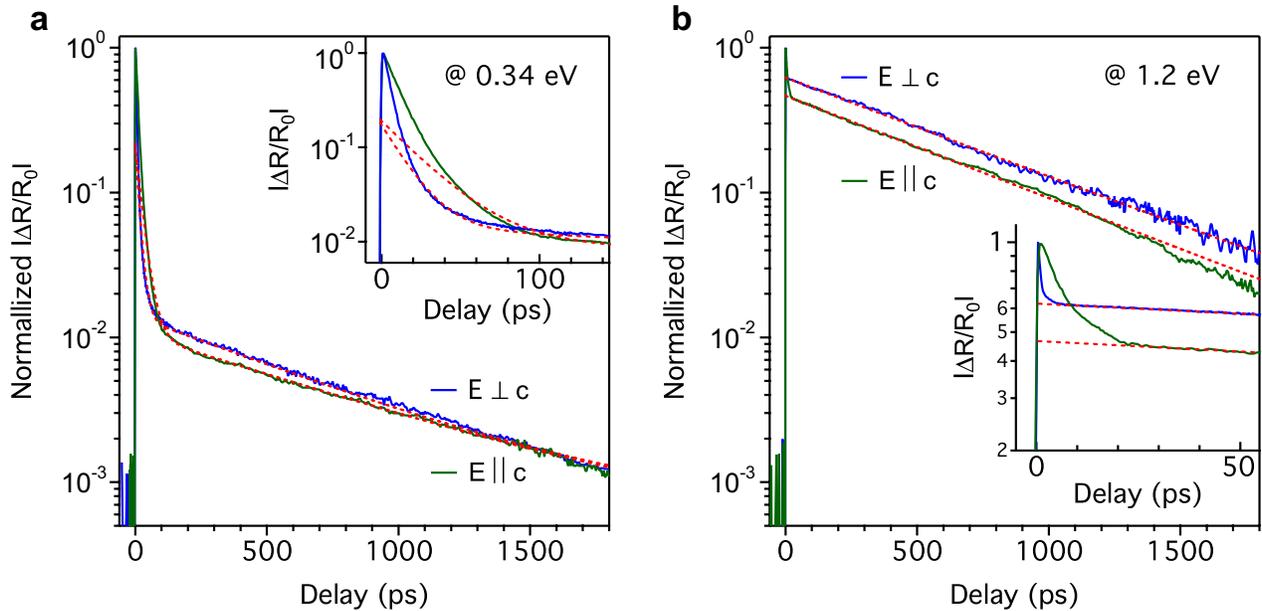

**Supplementary Figure 6 | Polarization and energy dependent carrier dynamics.** Representative polarization resolved transient reflectance (TR) $\Delta R/R_0$ traces of Te samples around the fundamental band-edge **a** and higher energy transition region **b**. Dashed red lines are the model fits of each corresponding data over a long delay range. Insets show zoom-in view, showing that the model does not cover early rapid carrier thermalization regime.

## Supplementary Note 5: Modeling transient reflectance spectroscopy (TRS) data

For simplicity, our model of the transient reflectance response of Tellurium is based on the dominant contribution of band-filling due to interband photoexcitation of carriers. We constrain our calculations by considering the dynamics near the fundamental gap at H-point of bulk Brillouin zone of Te. The photoinduced band filling effect alters the dielectric response of Te, i.e., a fractional change in the complex refractive index ($\tilde{n} = n + ik$), and causes a change in absorption and reflectance of the samples.

Since the imaginary component of the complex refractive index, i.e., $k$, relates to the absorption coefficient $\alpha$ as: $k = \frac{\lambda}{4\pi} \cdot \alpha$, we first begin the model by calculating the photoinduced change in the absorption coefficient $\Delta\alpha$. The absorption coefficient $\alpha$ of semiconductors with parabolic bands can be described by the following expression:[15, 16]

$$\alpha(E, N_h, N_e, T) = \frac{C_\alpha}{E} \int_0^{E-E_g} \rho_c(E') \cdot \rho_v(E' - E) \cdot \left[ f_h(E - E_g - E', N_h, T) - f_e(E', N_e, T) \right] dE' \quad (4)$$

Where $\rho_c$ and $\rho_v$ denote density of states for the conduction and valence bands, $N_h$ and $N_e$ are the respective carrier densities for holes and electrons, $f_h$ and $f_e$ are the appropriate Fermi-Dirac distributions, $T$ is the carrier temperature, and $C_\alpha$ is a constant factor fit to the data. Given the p-doped nature of our samples, we consider the total hole population $N_h$ as a sum of doped holes $N_d$ and photoexcited holes $\Delta N_h$, i.e., $N_h = N_d + \Delta N_h$. The electron population $N_e$ is taken to be entirely photoexcited, i.e., $N_e = \Delta N_e$, and the photoexcited electron and hole densities are taken to be equal, i.e., $\Delta N_e = \Delta N_h$, on the basis of charge neutrality. The carrier temperature $T$ is assumed to be equal for electron and holes. Now we calculate the photoinduced absorption coefficient $\Delta\alpha$ using relevant parameters, which are known for our samples (see Supplementary Table 1).

Once the photoinduced modulation in the absorption is calculated, we can derive the modulation of the real part of index of refraction, $\Delta n$, using the Kramers-Kronig relation:

$$\Delta n(E, N_h, N_e, T) = \frac{\hbar c}{\pi} \int_0^\infty \frac{\Delta\alpha(E', N_h, N_e, T)}{(E')^2 - E^2} dE' \quad (5)$$

where $\Delta\alpha$ is calculated from Eq. (4). Note that although the upper bound of the integral is infinite, $\Delta\alpha$ rapidly approaches zero above the band-edge and thus our theoretical description of the absorption need not include the high-energy regime.

Once we derive both the real part of refractive index and the absorption coefficient, we can connect to these parameters to evaluate the reflectance of the sample. Assuming normal incidence of the probe beam on the sample, which is the case in our experiment, the reflectance of the sample without excitation is given by:

$$R = \frac{(n-1)^2 + k^2}{(n-1)^2 - k^2} \qquad (6)$$

The fractional change of reflectance induced by pump excitation (normalized by initial reflectance $R_0$ before exciting the sample) can be expressed (in first-order expansion) as following:

$$\frac{\Delta R}{R_0} \cong \left[\left(\frac{8 n_o k_o}{((n_o+1)^2 + k_o^2)^2}\right) \cdot \Delta k(E, N_h, N_e, T)\right] \\ - \left[\left(\frac{4(1 + k_o^2 - n_o^2)}{((n_o+1)^2 + k_o^2)^2}\right) \cdot \Delta n(E, N_h, N_e, T)\right] \qquad (7)$$

Where $n_0$ and $k_0$ are the background values of the real and imaginary components of the complex index of refraction, respectively. We take $n_0$, for light polarized perpendicular to the c-axis, to be constant at the average value of 4.95 for our probe tuning range, based on measurements of single-crystal Tellurium.[17] We calculate $k_0 = (\lambda/4\pi) \cdot \alpha$ from Eq. (4) using background estimates for carrier density. All basic material parameters are tabulated in Supplementary Table 1.

We fit theoretical lineshapes from Eq. 7 to our experimental transient reflectance spectra by parameterizing the density of doped holes, density of photoexcited carriers, carrier temperature, band-gap energy, and an overall scaling factor. We numerically optimize these parameters to minimize the sum squared error for each spectrum taken at distinct delay times, keeping the time-independent parameters consistent. The resulting fit parameters are tabulated in Supplementary Table 2.

Supplementary Table 1: Basic material parameters of Te used for modeling.

| Parameters | Values |
| --- | --- |
| Electron effective mass*, $m_e^{*}$[4] | 0.091 $m_e$ |
| Hole effective mass*, $m_h^{*}$[4] | 0.137 $m_e$ |
| Refractive index, $n_0$[17] | 4.95 |

*Geometric mean accounting for anisotropy.

Supplementary Table 2: Band parameters of Te extracted from modeling TR spectra.

| Parameters | @ 10 ps delay | @ 30 ps delay | @ 60 ps delay |
|---|---|---|---|
| Carrier temperature, $T_e$ | 84 K | 67 K | 47 K |
| Photoexcited carrier density, $\Delta N_{eh}$ | $1.0 \times 10^{18}$ cm$^{-3}$ | $5.2 \times 10^{17}$ cm$^{-3}$ | $2.9 \times 10^{17}$ cm$^{-3}$ |
| Quasi-Fermi Energy (holes), $E_{fh}$ | 0.031 eV | 0.027 eV | 0.026 meV |
| Fundamental band gap, $E_g$ | 0.319 eV | 0.319 eV | 0.319 eV |
| Doping Density, $N_d$ | $1.8 \times 10^{18}$ cm$^{-3}$ | $1.8 \times 10^{18}$ cm$^{-3}$ | $1.8 \times 10^{18}$ cm$^{-3}$ |

## Supplementary Note 6: Polarization anisotropy of optical reflectance in Te

The ground-state optical transitions can be studied by measuring the linear optical absorption or reflectance of the samples. Such measurements are sometimes quite challenging in substrate (high absorbing substrates) supported ultrathin samples due to dominant contributions from the substrate. Linear dichroism, i.e., polarization anisotropy of optical absorption or reflectance, otherwise, can be very useful to reduce the substrate contributions and enhance the sensitivity of detecting optical response of the sample, particularly in anisotropic crystals. Here, we demonstrate that it is indeed possible to observe band-edge absorption features in a typical reflectance measurement of Te nanosheet samples supported on ~ 350 nm SiO$_2$/Si substrate. We observe the response of clearly distinguishable ground-state optical transitions, which are otherwise hardly visible from the reflectance spectra of the sample with substrate.

Supplementary Figure 7 displays probe photon energy dependent optical reflectance anisotropy (RA) or polarization anisotropy of Te nanosheets measured at 10 K. The RA is deduced by measuring the difference in reflectance for light normally incident and linearly polarized along two orthogonal directions, i.e., parallel $E \parallel c$ and perpendicular $E \perp c$ to the c-axis of the Te crystal. The normalized RA is expressed by: $RA = (R_\parallel - R_\perp)/(R_\parallel + R_\perp)$, where $R_\parallel$ and $R_\perp$ denote the reflectance along $E \parallel c$ and $E \perp c$, respectively. The individual reflectance spectra of each polarization are also shown on the top. The anisotropy is particularly pronounced around the band-edge regions, which are shown by thick grey lines and also indicated by particular transitions in the band structure. First minimum coincides with the $H_4 \to H_6^{CB}$ transition between the uppermost $H_4$ VB to the lowermost $H_6$ CBs, i.e., $E_1 = E_1^\perp$ and $E_1^\parallel$. The next maximum matches with the $H_5 \to H_6^{CB}$ transition between the spin-split lower $H_5$ VB to the $H_6$ CBs, i.e., $E_2 = E_2^{\parallel}$. The positive peak of the RA is due to 180° phase shift of this transition, which is dipole forbidden for $E \perp c$. Weak maxima around the other two higher energy transitions $E_3 = E_3^{\perp,\parallel}$ and $E_4 = E_4^{\perp,\parallel}$ can also be recognized in the spectra. Overall, these RA features are consistent with the features observed in transient reflectance spectra and further supports our estimations of ground-state optical transitions in our Te samples.

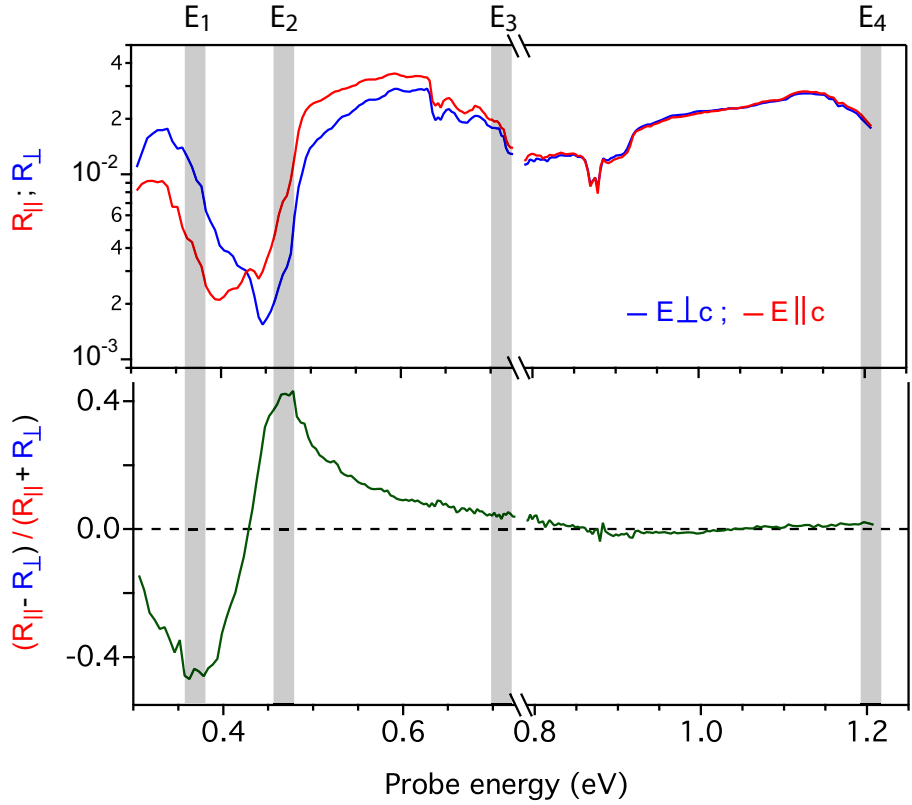

**Supplementary Figure 7 | Polarization anisotropy near the band-edge of Te nanosheets.** The polarization anisotropy (bottom plot) is calculated from the reflectance spectra (shown on the top) recorded for both $E \perp c$ and $E \parallel c$ polarizations. The anisotropy peaks near the band-edges of different bands around H-point of the Brillouin zone. Respective transition energies are also indicated with transparent grey thick lines. These transition energies are consistent with ground-state transitions extracted from the TR spectra (see main text).

## Supplementary Note 7: Coherent longitudinal acoustic phonon in Te nanosheets

Coherent longitudinal acoustic phonons (CLAP) are propagating ultrasonic strain waves in solid crystals, which can be excited efficiently at the surface by ultrafast laser pulses through transient lattice deformation.[18-21] Since the strain wave propagation modifies the local dielectric properties of the materials, the induced oscillatory behavior of dielectric properties can be probed in the time-domain by various pump-probe spectroscopic techniques.[18, 22]

Supplementary Figure 8 displays probe wavelength dependent TR response measured following 1.51 eV pump excitation of the Te nanosheet samples and probing at different photon energies (wavelengths). Following initial ultrafast transients, the TRS signal oscillates in time and gradually decays because of interference between light waves reflected off the surface and light waves reflected off the propagating strain wave. The TR signal, therefore, contains both the electronic response of the material as well as the oscillatory interference signal due to the CLAP generated in the sample. In order to extract information about the oscillatory signal, each trace is fitted with following empirical function:

$$\frac{\Delta R}{R_0}(t) = A + \left(B \cdot e^{-\frac{t}{\tau_1}}\right) + \left(C \cdot e^{-\frac{t}{\tau_2}}\right) + \left\{D \cdot \cos\left(\frac{2\pi t}{T} + \varphi\right) \cdot e^{-\frac{t}{\tau_{osc}}}\right\}, \qquad (7)$$

where first three terms fit the non-oscillatory electronic response including bi-exponential (initial fast $\tau_1$ followed by slower $\tau_2$) response and last one fits the slowly decaying oscillatory signal. Coefficients $B, C, D$ are amplitude of each decaying components and $A$ is the overall background signal. Least square fitting method is applied to each wavelength dependent trace, as shown by black line superimposed on each trace, and useful parameters such as the velocity of the CLAP are extracted as fit parameters. An average amplitude of the oscillation is of the order of $10^{-4}$ and damping constant $\tau_{osc}$ varies from 50-100 ps. Interestingly, the oscillation period $D$ remains unchanged for each probe wavelength. This behavior suggests that standing wave condition of CLAP propagation within the film is fulfilled, which is expected in ultrathin samples. The period of the oscillation is determined by film thickness $d$:

$$T = \frac{2 \cdot d}{v_S}, \qquad (8)$$

where $v_S$ is the velocity of CLAP (sound velocity), which depends on the polarization of probing laser pulse. An average oscillation period of $26.3 \pm 0.1$ ps estimates the average velocity $v_0$ of the LA phonon (sound velocity) of ~ 1800 m/s in our ultrathin ($d \sim$ 24 nm) Te samples. The estimated velocity $v_0$ is lower than both the sound velocity along parallel (3400 m/s) and perpendicular to c-axis (2290 m/s), but higher than the velocity along xy-plane or shear plane (1390 m/s),[23] suggesting quasi-shear wave propagation in our sample.[24]

It is important to discuss the physical origin of the acoustic phonon generation in our Te samples. There are mainly three possible mechanism of stress generation upon photoexcitation: (1) thermoelastic stress ($\sigma_{TH}$), (2) electron-acoustic deformation potential stress, $\sigma_{DP}$ and (3) inverse piezoelectric stress, $\sigma_{IP}$. The photoinduced $\sigma_{TH}$ due to rapid lattice heating by electron-phonon coupling can be estimated by following the standard model,[19] $\sigma_{TH} = -3 \cdot \beta \cdot B \cdot \Delta T$, where $\beta$ is the linear thermal expansion coefficient, $B$ is the bulk modulus and $\Delta T$ is the maximum temperature increased by pump pulse. Due to anisotropic crystal, Te has negative expansion coefficient parallel to c-axis, i.e., $\beta = -5\times10^{-6}\ K^{-1}$, and positive coefficient perpendicular to c-axis, i.e., $\beta = 5\times10^{-6}\ K^{-1}$. The bulk modulus in Te is $B = 19$ GPa.[25] It is known that electron temperature immediately after pump excitation of Te sample is extremely high due to excitation of hot carriers, which thermalizes to lattice temperature within a ps through electron-optic-phonon coupling.[26] The lattice temperature increases only few K given the low excitation density used in our low-temperature (10 K) experiment. Therefore, we can safely assume an upper bound increased lattice temperature of $\Delta T = 20$ K within 10 ps. Now, using all these parameters the photoinduced thermoelastic stress $\sigma_{TH}$ can be estimated to be $|\sigma_{TH}| = 0.006$ GPa, which will be even smaller at later delay times. Next, the electron-acoustic deformation potential stress $\sigma_{DP}$ can be estimated by:[19] $\sigma_{DP} = -a_{e-ac} \cdot \Delta N \approx -0.0003$ GPa, with $a_{e-ac} = -8.5$ eV along c-axis[27] and with photoinduced carrier density of $\Delta N \sim 2\times10^{23}$ m$^{-3}$. The stress perpendicular to c-axis increases to $\sigma_{DP} \approx 0.001$ GPa, with $a_{e-ac} = 35$ eV.[27] If we estimate the band gap change from these stress values,[28] these stress values are too small to induce tens of meV band gap shifts that we observed in our samples. Therefore, we argue that the piezoelectric strain induced by inverse piezoelectric effect (IPE) is the most likely mechanism of strain generation in our samples. Based on experimental observations, piezoelectric shear strain is dominant in our samples.

With the simple theoretical expression,[29, 30] $\Delta E_{split} \approx 4 \cdot \Psi_d \cdot \mathcal{E}_{xy}$, where $\Delta E_{split}$ is the band splitting or lifting, $\Psi_d$ is the shear deformation potential and $\mathcal{E}_{xy}$ is the shear strain, it is possible to estimate the shear strain value of the materials if we know the shear deformation potential and the resulting strain induced band splitting. The band splitting of $\Delta E_{split} = \Delta E_{lift} \sim 20$ meV is known from the TRS measurements at low temperature but the shear deformation potential is not known yet to the best of our knowledge. The typical value of deformation potential found in the literature is ~5 eV, which is estimated by the shift of energy gap in Te with dilation,[31] If we assume $\Psi_d \sim 5$ eV in our case, the required shear strain turns out to be on the order of ~ 0.1 %, which is substantially lower than the value estimated by our ab initio DFT band structure calculations, i.e., ~2 − 3 %. Since the shear deformation potential is fundamentally different and complex, it is hard to predict the exact value of shear strain using this simple expression. However, considering required maximum shear strain of ~ 2 % to induce 20 meV band splitting, we can estimate the shear deformation potential in Te to be 0.25 eV. This simple estimation of photoinduced shear strain and shear deformation potential in Te will be very helpful for future studies since photoexcitation can be an alternative and non-destructive means for inducing exotic topological phases in Te.

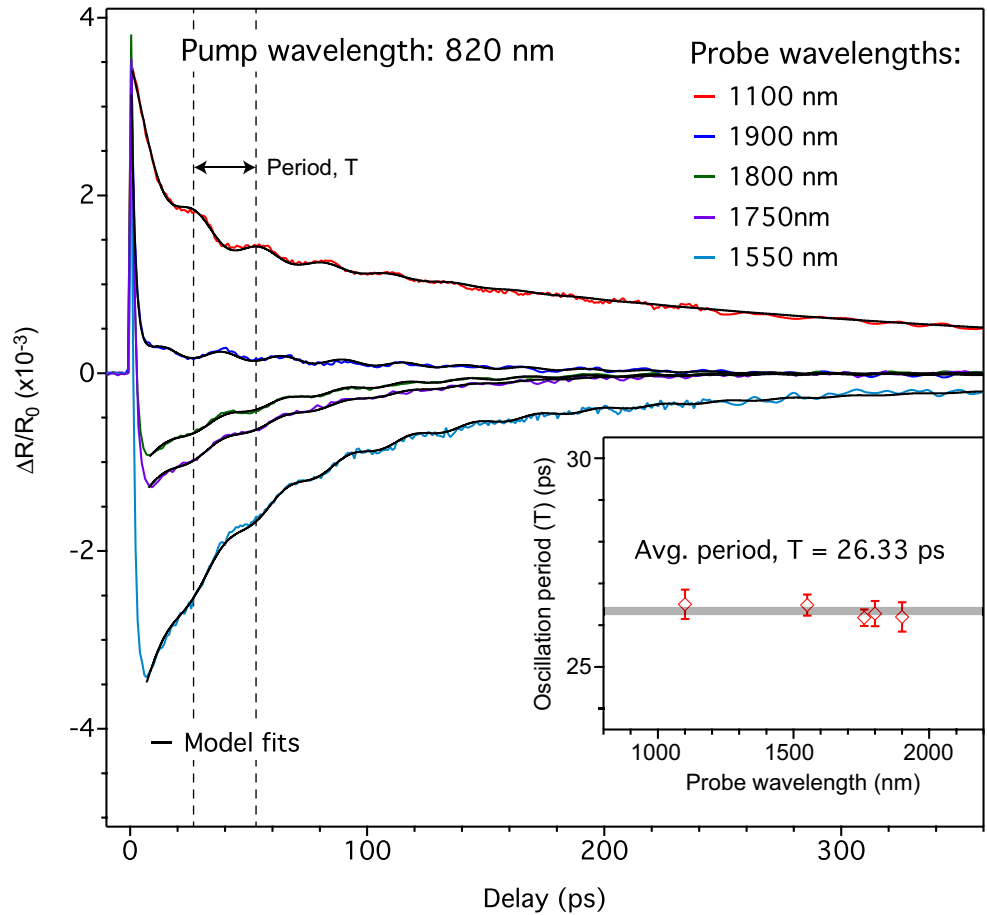

**Supplementary Figure 8 | Coherent longitudinal acoustic phonon in Te nanosheets.** Transient reflectance signal (TRS) of Te samples measured at different probe energies following 820 nm pump excitation at room temperature. Probe beam is polarized off-axis to ab-plane as well as c-axis but fixed for each wavelength. The TRS traces display weak oscillations embedded on a slowly decaying electronic signal. The oscillatory signal is caused by interference between the probe laser beam and photoinduced strain pulse propagation in the sample. Due to ultrathin samples, standing wave condition is fulfilled, which satisfy the linear dependence of oscillation period with film thickness, but does not depend on wavelength of the probe beam (see inset). This condition allows to estimate thickness knowing the acoustic wave velocity or vice versa through: $T = 2 \cdot d/v$, where $d$ is sample thickness, $T$ is the oscillation period, $v$ is the velocity of longitudinal acoustic phonon, i.e., sound velocity, in Te. An average period of the oscillations is estimated to be 26.33 ps, which estimates the sound velocity of $\sim$ 1800 m/s in our 24 nm thick Te sample.

# Supplementary References